\pdfoutput=1
\documentclass[conference]{IEEEtran}
\IEEEoverridecommandlockouts
\usepackage{cite}
\usepackage{amsmath,amssymb,amsfonts}
\usepackage{algorithmic}
\usepackage{graphicx}
\usepackage{textcomp}
\usepackage{xcolor}
\usepackage{authblk}

\usepackage{listings}
\usepackage{enumitem}
\newlist{steps}{enumerate}{1}
\setlist[steps, 1]{label = Step \arabic*:}
\usepackage{multirow}
\lstdefinestyle{CStyle}{
    commentstyle=\color{blue},
    keywordstyle=\color{magenta},
    numberstyle=\tiny,
    basicstyle=\scriptsize,
    breakatwhitespace=false,         
    breaklines=true,                 
    captionpos=b,                    
    keepspaces=true,                 
    numbers=left,                    
    numbersep=5pt,                  
    showspaces=false,                
    showstringspaces=false,
    showtabs=false,                  
    tabsize=2,
    language=C
}

\lstdefinestyle{aa}{
basicstyle=\ttfamily\scriptsize,
frame=single
}

\lstdefinestyle{bb}{
basicstyle=\scriptsize,
frame=single
}

\usepackage{mathptmx} 

\usepackage[normalem]{ulem}
\usepackage[hyphens]{url}
\usepackage[final]{microtype}
\usepackage{subfigure}

\usepackage[ruled,vlined]{algorithm2e}

\SetCommentSty{mycommfont}

\usepackage{subfig}

\usepackage{pifont}
\newcommand{\cmark}{\ding{51}}%
\newcommand{\xmark}{\ding{55}}%

\def\BibTeX{{\rm B\kern-.05em{\sc i\kern-.025em b}\kern-.08em
    T\kern-.1667em\lower.7ex\hbox{E}\kern-.125emX}}

\begin{document}
\title{Adversarial Prefetch: New Cross-Core Cache Side Channel Attacks} 

\author[1]{Yanan Guo}
\author[ \hspace{-0.7ex}]{Andrew Zigerelli}
\author[2]{Youtao Zhang}
\author[1]{Jun Yang}
\affil[1]{Electrical and Computer Engineering Department, University of Pittsburgh}
\affil[2]{Department of Computer Science, University of Pittsburgh}
\affil[ ]{yag45@pitt.edu,  zhangyt@cs.pitt.edu, juy9@pitt.edu}

\maketitle



\begin{abstract}

Modern x86 processors have many prefetch instructions that can be used by programmers to boost performance. However, these instructions may also cause security problems. In particular, we found that on Intel processors, there are two security flaws in the implementation of \texttt{PREFETCHW}, an instruction for accelerating future writes. First, this instruction can execute on data with read-only permission. Second, the execution time of this instruction leaks the current coherence state of the target data. 

Based on these two design issues, we build two cross-core private cache attacks that work with both inclusive and non-inclusive LLCs, named Prefetch+Reload and Prefetch+Prefetch. We demonstrate the significance of our attacks in different scenarios. First, in the covert channel case, Prefetch+Reload and Prefetch+Prefetch achieve 782 KB/s and 822 KB/s channel capacities, when using only one shared cache line between the sender and receiver, the largest-to-date single-line capacities for CPU cache covert channels. Further, in the side channel case, our attacks can monitor the access pattern of the victim on the same processor, with almost zero error rate. We show that they can be used to leak private information of real-world applications such as cryptographic keys.
Finally, our attacks can be used in transient execution attacks in order to leak more secrets within the transient window than prior work. From the experimental results, our attacks allow leaking about 2 times as many secret bytes, compared to Flush+Reload, which is widely used in transient execution attacks.

\end{abstract}

\begin{IEEEkeywords}
cache security, side channel attacks
\end{IEEEkeywords}

\section{Introduction}
\label{sec:intro}

Modern processors often feature many microarchitectural structures that are shared among applications. Although such resource sharing enables significant performance benefits, it also gives adversaries the potential to build powerful covert channel and side channel attacks. 
When an application runs on such hardware, its execution may cause various state changes to these shared microarchitectural structures, which can be observed by an attacker on the same platform. Through repeated observations, the attacker can derive the application's private information related to the state changes, bypassing sandboxes and traditional privilege boundaries.
Cache timing covert channel and side channel attacks, or cache attacks for short, are extremely potent~\cite{fr, ff, l1prime, pp, lru1, lru2, coherence_attack, template, aes_attack, profit, evict_time, cache_attack1,cache_attack2,cache_attack3,cache_attack4,cache_attack5,cache_attack6,cache_attack7,cache_attack8,cache_attack9,cache_attack10,cache_attack11,occupacy,prime+abort,everyday_app}. They are especially powerful primitives used in the more recently discovered transient execution attacks~\cite{spectre, meltdown, zombie, ridl, microscope, lvi, netspectre,checkmate,cacheout,smotherspectre,spectre_asplos21}. Different cache behaviors, such as hits and misses create significant timing differences to the execution of an instruction. Attackers can use these timing variances to stealthily transfer data (in the covert channel case) or infer some secrets from a victim (in the side channel case) such as cryptographic keys.

A critical step in most cache attacks is \textit{evicting the victim's data} from a cache level. 
Based on how the attacker evicts the victim's data, most cache attacks can be classified into \textit{flush-based} attacks~\cite{fr, ff, coherence_attack, coherence_attack_amd} and \textit{conflict-based} attacks~\cite{pp,ps,template}. Flush-based attacks usually assume data sharing between the  attacker and victim. Thus, the attacker directly performs \texttt{CLFLUSH} on the victim's data to evict it from all cache levels. For conflict-based attacks, the attacker instead achieves the eviction by constructing set conflicts, i.e., the attacker fills the cache set (that the victim's data occupies) with his own data.
Many secure cache designs have been proposed to defend cache attacks. For example, flush-based attacks can be prevented by modifying \texttt{CLFLUSH} (to make it a privileged instruction), as suggested in prior work~\cite{fr, ceaser-s, scattercache, cache3}. Conflict-based attacks can be defended by stopping/limiting attackers from discovering congruent addresses~\cite{ceaser, ceaser-s, scattercache}. Thus, in this work we present new cache eviction methods to enable practical cache attacks.

\begin{figure*}[!thb]\centering \begin{minipage}[b]{\textwidth}
\centering
\includegraphics[width=0.99\columnwidth]{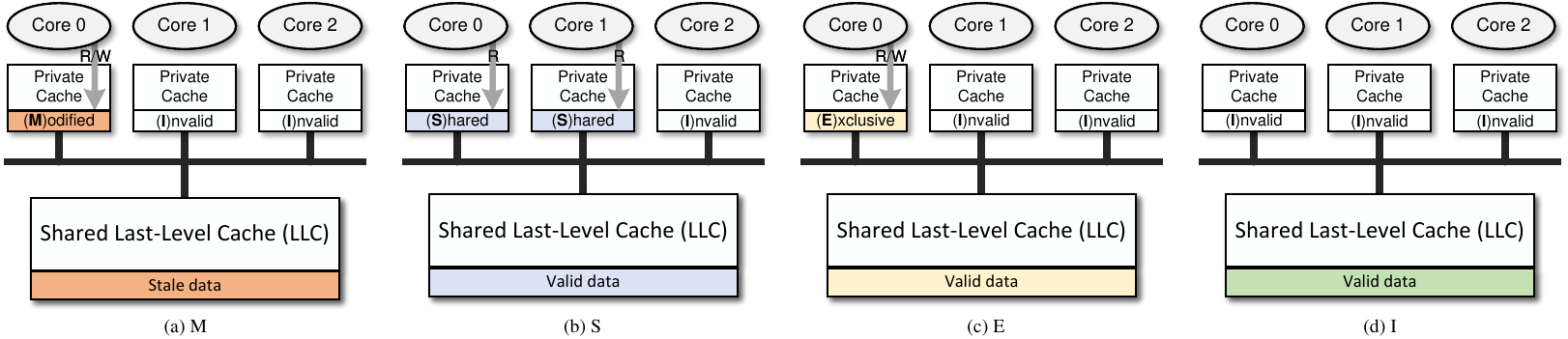}
\end{minipage} 
\caption{The four possible states of a private cache line, when using the MESI protocol.} \label{mesi} 
\end{figure*}


\texttt{PREFETCHW} is an x86 prefetch instruction introduced in 2000. It is now available on all Intel Xeon Scalable processors and recent Core processors (since Broadwell). According to the technology manual~\cite{intel_optimization_manual}, the function of this instruction is to prepare data for \textit{future writes}. It is different from other prefetch instructions (e.g., \texttt{PREFETCHT0}) which only move the target cache line closer to the CPU core (i.e., to a higher cache level) to get ready for future accesses. \texttt{PREFETCHW} moves the cache line to the requesting core's L1 data cache (L1D cache), as well as sets the coherence state of the cache line to be \textit{Modified}. This can accelerate future write operations from this requesting core, because a cache line in Modified state indicates that 1) the current private cache has exclusive ownership of this cache line, meaning a write operation on this cache line can be directly served by the private cache, and 2) this cache line is already marked as dirty, so the flag (i.e., the dirty bit) does not need to be changed when serving a write operation. 
For correctness, setting the coherence state of a cache line to Modified causes all copies of this cache line in other cores' private caches to be invalidated~\cite{amd_mesi, intel_mesi}.  

In this work, we make two important observations regarding \texttt{PREFETCHW} on Intel processors. First, although its purpose is to accelerate future writes, \texttt{PREFETCHW} works on data with \textit{read-only} permission. Second, the execution time of \texttt{PREFETCHW} is related to the current coherence state of the target data. With the first observation, an attacker on a different core than the victim can use \texttt{PREFETCHW} on the shared data between the attacker and the victim (which is usually read-only~\cite{fr}), to evict this data from the victim's private cache. In addition, the second observation means that the attacker can time the execution of \texttt{PREFETCHW} on the shared data between the attacker and victim to learn the coherence state changes of this data, which could be related to the victim's cache accesses. 



Based on these two observations, we first propose two covert channel attacks: Prefetch+Load and Prefetch+Prefetch. In Prefetch+Load, the sender transmits a bit by prefetching (with \texttt{PREFETCHW}) the shared data between the sender and receiver (for ``1'') or not prefetching (for ``0''). The receiver (on a different core) receives the bit by loading this data and timing the load to determine if it is a local private cache hit (for ``0'') or a remote private cache hit (for ``1''). In Prefetch+Prefetch, the sender transmits a bit by loading (or not) the shared data, and the receiver receives the bit by prefetching the data and timing the prefetch instruction to determine whether the sender loaded. We show that our prefetch-based channels have very high capacities: on our Kaby Lake processor, when only using one shared cache line between the sender and receiver, the capacities are 840KB/s for Prefetch+Load, and 822KB/s for Prefetch+Prefetch, which are \textit{the highest single-line capacities among all existing CPU cache covert channels}. 

We then modify the covert channel attacks and build the Prefetch+\textbf{Re}load and Prefetch+Prefetch side channel attacks, which can be used to leak the victim's access patterns, similar to previous cache attacks (e.g.,~\cite{fr, ff, pp, lru1, lru2, coherence_attack, l1prime}). Prefetch+Prefetch can be directly used as a side channel attack by letting the victim be the sender, and the attacker be the receiver, since in this attack the sender transmits signals by accessing (or not) the shared data. However, in Prefetch+Load, the sender is sending signals by prefetching (or not), which is unlikely a side channel. Thus, we modify it and build Prefetch+Reload, where the attacker owns two threads running on different cores. The attacker first uses one thread to prefetch and waits for the victim's possible access, and then reloads using the other thread. When the attacker reloads, he will get a remote private cache hit if the victim accessed this data; otherwise he will get a last level cache (LLC) hit. Then, the attacker can determine the victim's behavior using timing information: a remote private cache hit and an LLC hit take different amounts of time to finish.  
We show that our attacks can be deployed on Intel processors to leak secrets from real-world applications, and that they can be used in transient execution attacks, making those attacks faster (and more potent) than before. 
To the best of our knowledge, our prefetch-based attacks are the first cross-core \textit{private cache} side channel attacks that can work with both inclusive and non-inclusive LLCs: in our attacks, the victim's data is only evicted from the private cache but never the LLC.

In this paper, we make the following contributions:
\begin{itemize}
    \item We discover two severe security vulnerabilities in the implementation of \texttt{PREFETCHW}.
    \item We present a new cache eviction method, as well as two new cross-core cache covert channels and side channels, using \texttt{PREFETCHW}.
    \item We evaluate the proposed prefetch-based covert channels and side channels on multiple desktop and server processors. The experimental results show that 1) our covert channels are faster than most existing cache covert channels and 2) our side channel attacks can leak information from daily applications with high temporal resolution.
\end{itemize}

We have disclosed the security vulnerabilities we found in this paper to Intel. The source code of our attacks can be found at \textcolor{blue}{\url{https://github.com/PittECEArch/AdversarialPrefetch}}.


\section{Background}
\label{sec:background}

\subsection{CPU Cache Architecture and Coherence Protocol}
\label{cache}
\noindent
\textbf{Cache architecture.} 
Most CPU caches on modern x86 processors are divided into L1, L2, and L3. The L1 and L2 caches are very fast but relatively small. Typically, they are organized separately for each CPU core, and are thus often referred to as private caches. In contrast, the L3 cache, also known as last-level cache (LLC), is a larger but slower cache, shared among CPU cores.

Caches operate on fixed-size (e.g., 64 bytes) data blocks  called cache lines.
Additionally, caches are usually set-associative: a cache is organized into multiple cache sets. Every cache set consists of multiple equivalent cache ways, and each of them can store one cache line. The address bits of a cache line determine which cache set that this line is mapped to. Most LLCs in Intel processors are inclusive, meaning that data present in private caches are necessarily also present in the LLC (and conversely, data not in the LLC are not in the private caches). However, recent Intel server processors (e.g., Intel Xeon Scalable processors~\cite{skylake-sp, non-inclusive-attack}) use non-inclusive LLCs, i.e., cache lines in private caches may not be present in the LLC. For non-inclusive LLCs, a separate directory structure is required for tracking the cache lines that are in the private caches but not in the LLC. 

When a CPU core performs a memory access request, it first checks whether the target cache line is present in its L1 or L2 cache. If present, the request results in a private cache hit; if not, it is a private cache miss and the core must further check the LLC (and the directory for a non-inclusive LLC). If the cache line is found, the request finishes and the data is sent to the CPU. If not, the cache forwards the request to the memory controller, which can read data from DRAM.


\noindent
\textbf{Cache coherence.} 
In multi-core systems, a cache line can be present in multiple private caches, due to data sharing. A cache coherence protocol\footnotemark  ~is required for maintaining data consistency among the copies of a cache line in different private caches: each private cache line is assigned a coherence state, and the LLC needs to track this state to prevent the use of stale data. For inclusive LLCs, the coherence states of private cache lines are stored together with the tag array in the LLC since all the private cache lines are also in the LLC. For non-inclusive LLCs, the directory structure mentioned earlier is used for storing the coherence states of cache lines that are in the private caches but not the LLC.

\footnotetext{In this paper, we only focus on the cache coherence inside a processor; this should not be confused with the coherence among sockets (processors)~\cite{coherence_attack_amd}.}

Most modern x86 processors use variants of the MESI coherence protocol~\cite{intel_mesi, amd_mesi}. In the rest of this section, we use inclusive cache as an example to introduce MESI. For non-inclusive caches, the protocol is essentially the same, except that a cache line in a private cache might not be present in the LLC. With MESI, there are four possible states of a private cache line:
\begin{itemize}

    \item \textit{\textbf{Modified (M)}}, in which the cache line is only present in one private cache and is dirty, i.e., the copy of this cache line in the LLC contains stale data (Figure~\ref{mesi}(a)). Additionally, when a private cache line is in M state, the current owner core has read/write permission for it.  
    \item \textit{\textbf{Shared (S)}}, in which the cache line is present in one or more private caches and is clean, i.e., the data of this cache line matches all other copies (both in other private caches and the LLC). The current core can only read this cache line (Figure~\ref{mesi}(b)).
    \item \textit{\textbf{Exclusive (E)}}, in which the cache line is only present in one private cache, and is clean (Figure~\ref{mesi}(c)). The current core can read/write this cache line; however, a write operation will change the state of this cache line to M.

    \item \textit{\textbf{Invalid (I)}}, in which the cache line is invalid, and thus the current core has neither read nor write permission for it (Figure~\ref{mesi}(d)).
\end{itemize}

With MESI, a memory request from a CPU core will sometimes 1) change the coherence state of the target cache line, and 2) take different amounts of time to finish, depending on the coherence state of the target cache line. 

\begin{figure}[!h] \centering
 \begin{minipage}[b]{0.99\columnwidth}
\centering
\includegraphics[width=0.99\columnwidth]{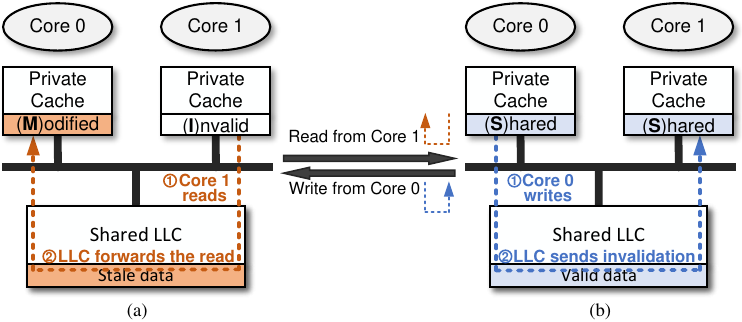}
\end{minipage} 
\caption{The illustration of cache coherence state changes. The state of a line changes from M (shown in (a)) to S (shown in (b)) when a CPU core is loading it; conversely, the state changes from S to M when a CPU core is writing it. Dashed lines shows the request path of the read/write operation. } \label{coherence_MtoS} 
\end{figure}

\noindent
\textbf{State transitions.} There are many different coherence state transitions, we only discuss the two scenarios related to our attacks. First, as shown in Figure~\ref{coherence_MtoS}(a), when a CPU core (core 1) is reading a cache line that is present in the LLC and the private cache of another core (core 0) in M state, this read request will first miss in its private cache and then search the LLC. Although this target cache line can be found in the LLC, its content is potentially stale. Thus, the LLC will fetch the data from the owner private cache (in core 0) that contains the up-to-date data of this cache line, change the coherence state of this cache line in that private cache (in core 0) to S, update the content of this cache line in the LLC, and then return the updated cache line to the requesting core (core 1) as well as fill its private cache with a copy of this cache line in S state. Thus, after serving this read request, the target cache line is present in two private caches, and is in S state in both caches, as shown in  Figure~\ref{coherence_MtoS}(b). This case is usually referred to as \textit{remote private cache hit}.

 \begin{figure}[!h] \centering
 \begin{minipage}[b]{0.99\columnwidth}
\centering
\includegraphics[width=0.99\columnwidth]{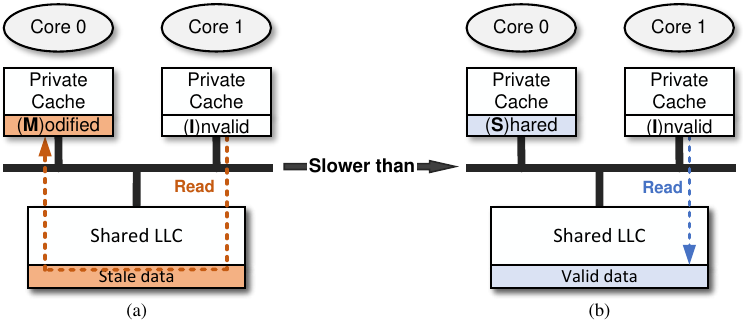}
\end{minipage} 
\caption{The illustration of an LLC access with the target cache line in M state (a), and S state (b). } \label{coherence_speed} 
\end{figure}

Second, as shown in Figure~\ref{coherence_MtoS}(b), when a CPU core (core 0) is trying to write a cache line that is in S state in its own private cache, this private cache (in core 0) needs to send request to the LLC to acquire write permission before it can serve this write operation. As a result, the LLC will send invalidation signal(s) to the other private cache(s) that the cache line is present in (in core 1), and then change the state of the cache line in the private cache of the requesting core (core 0) to M so that the requesting core can write this cache line in its private cache. Thus, after this write operation, the target cache line is only present in the requesting core's private cache, and is in M state, as shown in  Figure~\ref{coherence_MtoS}(a).

\noindent
\textbf{Timing difference.}
As one can observe in Figure~\ref{coherence_speed}, if a CPU core is reading a cache line that is not present in its private cache but is present in the LLC, 
the total latency it takes to finish this read request can be different when this cache line has different coherence states: a remote private cache hit is much slower than an LLC hit. 
When another core has a copy of this cache line in M state in its private cache, this request results in a remote private cache hit. As explained earlier, serving this request will require fetching data from the owner private cache. In contrast, when all the private cache copies of this cache line are in S state, the data of this cache line in the LLC is up-to-date. This means the LLC can serve this read request directly, resulting in an LLC hit. Due to these different execution paths, an LLC hit is much faster than a remote private cache hit.  
This has been observed by previous work~\cite{coherence_attack} and has been verified in our experiments.
On an Intel Core i7-6700 processor, an LLC hit takes less than 60 cycles to finish and a remote private cache hit takes about 90 cycles.

\subsection{Prefetch}
\label{pre_background}
Prefetch is a technique to boost performance by fetching data and placing them closer to the CPU core (e.g., from the LLC to L1 cache) before they are needed. Prefetch can be performed in two ways: 1) hardware prefetch, which is implemented in cache hardware and is transparent to users (e.g., the adjacent cache line prefetcher); 2) software prefetch, which needs to be explicitly done by the programmer/compiler. Recent x86 CPUs offer many different instructions for software prefetch, such as \texttt{PREFETCHT0}, \texttt{PREFETCHT1}, \texttt{PREFETCHT2}, \texttt{PREFETCHNTA}, and \texttt{PREFETCHW}.\footnote{Some CPU models (e.g., Intel Xeon Phi Processor 7200) use \texttt{PREFETCHWT1} instead of \texttt{PREFETCHW}.} These instructions are used to hint the processor that a memory location is very likely to be accessed soon~\cite{intel_optimization_manual}, then the processor will prefetch the corresponding data into certain level of cache, thereby accelerating future accesses to this data. Software prefetch is an important way to improve performance. For example, compilers sometimes inject prefetch instructions to accelerate for loops.

\subsection{Cache Side Channel Attacks}
\label{attack_back}
There are typically two types of cache attacks. The first type utilizes the contention on certain cache hardware (e.g., the ring interconnect~\cite{ring} or L1 cache ports~\cite{l1_port1, l1_port2}): the attacker passively monitors the latency of accessing this hardware resource to infer the victim's usage of it. Such attacks are usually referred to as contention based attacks or stateless attacks. 
The other type utilizes cache states: the attacker actively brings the cache line/cache set to a certain state, then lets the victim execute (which potentially changes the state), and later checks the state again to infer the victim's behavior.
Such attacks are often referred to as eviction based attacks or stateful attacks. In this overview, we focus on stateful attacks because they are more numerous, and our proposed attacks are stateful. We further divide stateful attacks into private cache attacks and LLC attacks, based on whether the attacker evicts the victim's data from the LLC during the attack.

\noindent
\textbf{Private cache attacks.} 
In private cache attacks, the attacker learns the victim's cache behavior by monitoring the state of the victim's data in the private cache. For example, in L1 Evict+Reload, the attacker evicts the victim's data (which is shared with the attacker) from the L1 cache to the L2 cache by building set conflicts, and waits for the victim's execution. Later the attacker accesses this data and times the access to determine it is in the L1 or L2 cache: if it is in the L1 cache, it means the victim accessed the data and brought it back to the L1 cache, otherwise the victim did not access.
Private cache attacks could have high-bandwidth since they do not create slow DRAM accesses. However, most private cache attacks require the attacker to be on the same physical core with the victim (e.g.,~\cite{lru1,evict_time}), and many of them further require SMT. This significantly limits the attacks, as cloud providers may allocate users to different cores and may disable SMT for security~\cite{dis_smt1,dis_smt2,dis_smt3}. 

The directory Prime+Probe attack~\cite{non-inclusive-attack} and its optimization, the directory Prime+Scope attack~\cite{ps}, are an exception: they are cross-core private cache attacks. On a processor with a non-inclusive LLC, the attacker can ``remotely'' evict the victim's data from the victim's private cache to the LLC (but not to DRAM) by building conflicts in the directory. 

\noindent
\textbf{LLC attacks.}
In LLC attacks (e.g., \cite{fr, llc_attack_dac, pp}), the attacker monitors the state of the victim's data in the LLC. The LLC is usually shared among CPU cores. Thus, different than private cache attacks, LLC attacks do not require the attacker to be on the same core as the victim. These attacks are considered more practical. 
However, DRAM accesses are usually involved in LLC attacks. To monitor the victim's access on the LLC data, the attacker needs to first evict the victim's data from the LLC to memory. 
For example, in Flush+Reload~\cite{fr}, the attacker uses \texttt{CLFLUSH} instruction to flush the victim's data from the LLC (and also the private caches), and later reloads this data and times this operation to determine whether the victim has brought this data back to the LLC.
Therefore, the bandwidths of LLC attacks are bottlenecked by DRAM latencies. 



\section{Characterizing Data Prefetching}
\label{characterization}
Among the prefetch instructions discussed in Section~\ref{pre_background}, \texttt{PREFETCHW} (or \texttt{PREFETCHWT1} on some CPU models) works slightly differently than the others. It not only brings the data close to the CPU core, but also changes the coherence state of the data: \texttt{PREFETCHW} places the target data cache line into the L1D cache\footnote{\texttt{PREFETCHW} can only be used on data but not instructions~\cite{3dnow,intel_optimization_manual}. Thus, the cache line will be brought into L1D cache.} and sets the coherence state of this cache line to M. According to the technology manual~\cite{3dnow, intel_optimization_manual}, the role of \texttt{PREFETCHW} is to \textit{accelerate future writes on the target cache line}. As explained in Section~\ref{cache}, the CPU core can directly write a cache line in its local L1 cache iff the state of this cache line is E/M. Thus, \texttt{PREFETCHW} pre-sets the coherence state of the target cache line to M so that future writes on this cache line will likely have an L1 hit. \texttt{PREFETCHW} sets the cache line state to M instead of E because writing a cache line in E state results in changing the state to M, and thus has higher latency than writing a cache line that is already in M state.

Most of the recent Intel desktop and server processors (since Broadwell) support \texttt{PREFETCHW}.
When used appropriately, it can significantly improve performance. However, we make two observations about \texttt{PREFETCHW} on Intel processors, which can be leveraged to create security vulnerabilities.

%

\vspace{0.1in}
\begin{itemize}[leftmargin=0pt ]
    \item[] \textbf{Observation 1} \textit{\texttt{PREFETCHW} successfully executes on data with read-only permission.}
\end{itemize} 
 \vspace{0.1in}

\begin{figure}[t]
\begin{minipage}[!b]{0.9\columnwidth}
\begin{lstlisting}[style=CStyle][t]
void* thread0 (void* addr_d0, int expt_idx){
    for(int i = 0; i < 1000000; i++){
        /* check the experiment index*/
        if(expt_idx == 0){
            /* execute prefetchw on d0*/
            prefetchw(addr_d0);}
        /*let thread1 execute 1 iteration*/
        wait_for_thread1(); 
    }}

void* thread1 (void* addr_d0){       
    for(int i = 0; i < 1000000; i++){
        /*let thread0 execute 1 iteration*/
        wait_for_thread0(); 
        int result = read_and_time(addr_d0);
    }}
    
    
int main() { 
    /* open and map a file as read-only*/
    int fd = open(FILE_NAME, O_RDONLY);
    int* addr_d0 = mmap(fd, PROT_READ, ...);
    
    /*pin thread0 on core0 and start thread0*/
    /*pin thread1 on core1 and start thread1*/
    ...
\end{lstlisting}
\captionof{lstlisting}{The code snippet for verifying Observation 1.}
\label{observation1_code}
\end{minipage}
\end{figure}

We observe this by monitoring the coherence state changes of the data, using timing information. Specifically, as shown in Listing~\ref{observation1_code}, we run a program with two threads ($thread_0$ and $thread_1$, both in one process), and pin them on different physical cores. We use \texttt{mmap}~\cite{mmap} to map part of a system file (e.g., glibc) as a read-only data block (in cache line size) in this program and name it $d_0$: both threads can only read $d_0$. If any thread tries to write $d_0$, it will trigger a segmentation fault. $thread_0$ and $thread_1$ both consist of a for loop with the same amount of iterations. In each iteration, $thread_0$ first executes, then waits for $thread_1$ to execute. After $thread_1$ finishes this iteration, they both move to the next iteration and repeat this procedure again. We use pthread mutex locking~\cite{pthread_lock} to ensure that in each iteration $thread_0$ and $thread_1$ execute sequentially (the implementation details of locking is omitted in Listing~\ref{observation1_code}). 

We run the code in Listing~\ref{observation1_code} twice: in the first experiment (i.e., expt\_idx = 0 in line 3), in each iteration of the for loop, $thread_0$ performs \texttt{PREFETCHW} on $d_0$, and then $thread_1$ loads $d_0$ as well as times the load. In the second experiment (i.e., expt\_idx = 1 in line 3), in each iteration $thread_0$ stays idle and then $thread_1$ still loads $d_0$ and times the load.

\begin{figure}[t]
\begin{minipage}{0.9\columnwidth}

\begin{lstlisting}[style=CStyle]
void* thread0 (void* addr_d0, int expt_idx){
    for(int i = 0; i < 1000000; i++){
        /* check the experiment index*/
        if(expt_idx == 0){
            read(addr_d0);}
        /*let thread1 execute 1 iteration*/
        wait_for_thread1() 
    }}
    
void* thread1 (void* addr_d0){       
    for(int i = 0; i < 1000000; i++){
        /*let thread0 execute 1 iteration*/
        wait_for_thread0();
        int t1 = rdtscp(); /* read time stamp*/
        prefetchw(addr_d0);
        int result = rdtscp()-t1;
    }}
    
int main() { 
    /* open and map a file as read-only*/
    int fd = open(FILE_NAME, O_RDONLY);
    int* addr_d0 = mmap(fd, PROT_READ, ...);
    
    /*pin thread0 on core0 and start thread0*/
    /*pin thread1 on core1 and start thread1*/
    ...
\end{lstlisting}
\captionof{lstlisting}{The code snippet for verifying Observation 2.}
\label{observation2_code}
\end{minipage}
\end{figure}

 \begin{figure}[!h] \centering
\vspace{0.1in}
\subfigure{
 \begin{minipage}[b]{0.48\columnwidth}
\centering
\includegraphics[width=\columnwidth]{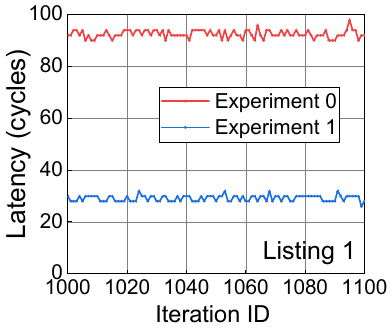}
\end{minipage}}
\subfigure{
 \begin{minipage}[b]{0.48\columnwidth}
\centering
\includegraphics[width=\columnwidth]{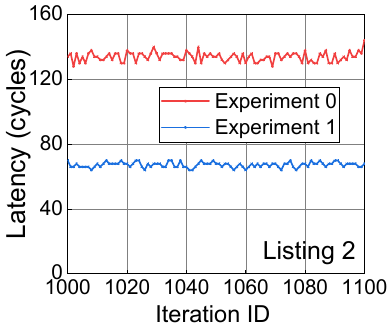}
\end{minipage}}
\caption{The timing measurement results in $thread_1$ of Listing~\ref{observation1_code} and Listing~\ref{observation2_code}. } \label{character1} 
\vspace{0.1in}
\end{figure}

Figure~\ref{character1} shows a segment of the timing results from $thread_1$ (in line 15) in both of the above experiments on an Intel Core i7-6700 processor. Note that we observe similar results on other Intel processors that support \texttt{PREFETCHW}. In experiment 0, $thread_0$ prefetches $d_0$ in each iteration, which causes $thread_1$ to take around 90 cycles to load $d_0$ after the prefetch. In contrast, in experiment 1, $thread_0$ stays idle, which causes $thread_1$  to take only around 30 cycles to load $d_0$. This timing difference infers that $d_0$ is in different states in the above two experiments. In experiment 0, every time when $thread_0$ prefetches, it will load $d_0$ to its private cache and set the coherence state of it to be M. According to MESI, explained in Section~\ref{cache}, this will invalidate the copy of $d_0$ in the private cache of $thread_1$ (if it exists). Therefore, when $thread_1$ later loads $d_0$, it will have a \textit{remote private cache hit} (see Figure~\ref{coherence_MtoS}). This load also changes the state of $d_0$ from M to S and fills a copy of it in the private cache of $thread_1$. Thus, the same cache behavior (i.e., invalidating the copy of $d_0$ in the private cache of $thread_1$) will happen when $thread_0$ prefetches in the next iteration. However, in experiment 1, since $thread_0$ is not prefetching, when $thread_1$ loads $d_0$, it will very likely have a \textit{local private cache hit}, which is much faster than a remote private cache hit (30 cycles\footnote{Due to the granularity of time stamp counters, this measured latency is in fact longer than the real private cache access latency.} vs. 90 cycles on the tested processor).

\noindent
\textbf{Rationale.}
We observe reliable cache state changes on read-only data when executing \texttt{PREFETCHW}, with a F-score of 1.0 ($n=1000000$). 
This indicates that Intel processors very unlikely perform a \textit{write permission check} when executing \texttt{PREFETCHW}.
This does not cause any error in the architecture level, because \texttt{PREFETCHW} only has microarchitectural effects: although it can get a cache line ready for future writes, if later the program without write permission for this cache line actually tries to write it, it will still trigger a fault and likely terminate the process. However, later we will show that allowing coherence-based cache invalidation (which should only happen upon writes) on read-only data cause significant security problems. This is because in cache attacks based on shared memory, the attacker can manipulate the coherence state of the shared data (which is usually read-only) between the victim and attacker to monitor the victim's access to this data.

\vspace{0.1in}
\begin{itemize}[leftmargin=0pt ]
    \item[] \textbf{Observation 2} \textit{The execution time of \texttt{PREFETCHW} is related to the coherence state of the target cache line.}
\end{itemize} 
\vspace{0.1in}

We observe this with the program shown in Listing~\ref{observation2_code}. We still use two threads pinned on different physical cores, and let them execute sequentially in each iteration of the for loop. Again, we run the program twice: in experiment 0 (expt\_idx = 0 in line 3), in each iteration, $thread_0$ loads $d_0$, and then $thread_1$ performs \texttt{PREFETCHW} on $d_0$ and times the prefetch. In experiment 1, $thread_0$ stays idle and $thread_1$ still prefetches and times the prefetch in each iteration.

Figure~\ref{character1} shows the execution time of \texttt{PREFETCHW} observed by $thread_1$ (in line 16) on our Intel Core i7-6700 processor in both experiments. In the first experiment, it always takes around 130 cycles for \texttt{PREFETCHW} to finish; however, in the second experiment it only takes around 70 cycles. This is because in the first experiment, after $thread_0$ loads $d_0$, the state of $d_0$ becomes S, and a copy of $d_0$ is filled into the private cache of $thread_0$ (see Figure~\ref{coherence_MtoS}). Then when $thread_1$ prefetches, it needs to change the state from S to M, which means it has to inform the LLC to invalidate the copy of $d_0$ in the private cache of $thread_0$. However, in the second experiment, since $thread_0$ stays idle, when $thread_1$ prefetches, $d_0$ is already in M state. Thus, in this case \texttt{PREFETCHW} does not cause any state change and finishes much faster.  

\noindent
\textbf{Affected processors.}
We have tested these two observations on many
Intel processors including the available 1st/2nd/3rd Generation Intel Xeon Salable Processors on AWS EC2, and five Intel desktop/server processors we own.  
As shown in Table~\ref{affected_proc}, Observation 1 is valid on all the tested processors, and Observation 2 is valid on most, excluding the Intel Xeon Platinum 8375C processor. On this processor, there is no difference on the execution time of  \texttt{PREFETCHW} when the target data is different coherence states: \texttt{PREFETCHW} always takes 70 to 80 cycles to finish, even when the target data is not already in M state.

In general, we believe that Observation 1 should be valid on all Intel processors that support \texttt{PREFETCHW}, and Observation 2 should be valid on most of them. Note that all 1st/2nd/3rd Generation Intel Xeon Scalable processors and most Intel Core i7/i9 processors (other than the early generations before Broadwell) support \texttt{PREFETCHW}. 

\begin{table}[!h]
\vspace{0.1in}
\footnotesize
\begin{center} \caption{The evaluated processors for the two observations.} \label{affected_proc} 
\setlength\arrayrulewidth{1.0pt}
\renewcommand{\arraystretch}{1.2}
    \setlength{\tabcolsep}{0.7mm}{ \begin{tabular}{lrrrr} 
    \hline
     Processor & Microarch. & LLC Type &Observ.1 & Observ.2 \\ \hline
     Intel Core i7-6700 & Skylake & Inclusive & \cmark & \cmark\\ 
     Intel Core i7-6800K & Skylake & Inclusive & \cmark & \cmark\\ 
     Intel Core i7-7700K & Kaby Lake & Inclusive & \cmark & \cmark\\ 
     Intel Core i9-10900X & Cascade Lake & Non-incl. & \cmark & \cmark\\ 
     Intel Xeon Silver 4114 & Skylake-SP & Non-incl. & \cmark & \cmark\\ 
     Intel Xeon Platinum 8151 & Skylake-SP & Non-incl.  & \cmark & \cmark\\ 
     Intel Xeon Platinum 8124M & Skylake-SP & Non-incl. & \cmark & \cmark\\ 
     Intel Xeon Platinum 8175M & Skylake-SP & Non-incl. & \cmark & \cmark\\ 
     Intel Xeon Platinum 8259CL & Skylake-SP & Non-incl. & \cmark & \cmark\\ 
     Intel Xeon Platinum 8275CL & Skylake-SP & Non-incl. & \cmark & \cmark \\ 
     Intel Xeon Platinum 8375C & Ice Lake & Non-incl. & \cmark & \textcolor{red}{\xmark}\\ 
     \hline
\end{tabular}} \end{center}  
\end{table}

\section{Prefetch-Based Covert Channel Attacks}
\label{covert}
Based on the observations in Section~\ref{characterization}, we build two cache covert channel attacks: Prefetch+Load and Prefetch+Prefetch. In this section, we first introduce the threat model, then discuss the details of each attack.
\subsection{Threat Model}
\label{threat_covert}
We assume that the two essential parties in the attack, the sender and receiver, are two unprivileged processes that are running on the same processor with multiple CPU cores. The sender and receiver can launch themselves on different physical cores (e.g., using \texttt{taskset}~\cite{taskset_real}). We also assume that the sender and receiver can share data; however, the shared data can be read-only (e.g., via shared library or page deduplication),\footnote{Page deduplication (a.k.a kernel same-page merging~\cite{ksm}) was originally created for virtual environments but is now included in OSs. Although many cloud providers no longer use it, it is usually still available in OSs~\cite{lru2}.} similar to previous attacks~\cite{coherence_attack, lru1,ff, fr,lru2}. In addition, the sender and receiver should agree on pre-defined channel protocols, including the synchronization, core allocation, data encoding, and error correction protocols. We do not have requirement on the LLC inclusivity; our attacks work with both inclusive and non-inclusive LLCs. We also do not require SMT; SMT can be turned off for security.


\subsection{Prefetch+Load Attack}
We build the first covert channel attack, Prefetch+Load, following Observation 1. Algorithm~\ref{alg1} shows the details of it. In this attack, the sender and receiver first agree on the shared cache line used to transmit information. Then in each iteration of the attack, the sender transmits a bit ``1'' by performing \texttt{PREFETCHW} on the shared cache line, or a bit ``0'' by idling. The receiver loads the same cache line and times the load to determine if it is a remote private cache hit or local private cache hit: the receiver receives a bit ``1'' when having a remote private cache hit, and otherwise receives a bit ``0''. 

Note that different than the experiments in Section~\ref{characterization}, the sender and receiver cannot synchronize using pthread mutex locking, since they do not belong to the same process. Thus, we let the sender and receiver synchronize the transmission using time stamp counters (TSCs), as done in prior covert channel attacks (e.g.,~\cite{lru1, coherence_attack, fr, ff, ps}).


\begin{algorithm}[t]
\scriptsize
\vspace{0.02in}
\SetKwFunction{sync}{sync\_with\_receiver}
\SetKwFunction{syncc}{sync\_with\_sender}
\textbf{line0}: the shared cache line between the sender and receiver\\
\textbf{message[n]}: the n-bit long message to transfer on the channel\\
\textbf{Th0}: the timing threshold for distinguishing local and remote private cache hit\\
\vspace{-0.02in}
--------------------------------------------------------------------------------------\\
\vspace{-0.02in}
\textbf{\scriptsize Sender Algorithm}\\
\vspace{-0.03in}
--------------------------------------------------------------------------------------\\
\tcp{Send 1 bit in each iteration.}
\For{\text{$i=0;$ $i < n;$ $i++$}}{
\sync{};\\
\eIf{$message[i] == 1$} {
        Prefetch line0;
    }
    {
        Do not prefetch;
    }
}

--------------------------------------------------------------------------------------\\
\vspace{-0.02in}
\textbf{\scriptsize Receiver Algorithm}\\
\vspace{-0.03in}
--------------------------------------------------------------------------------------\\
\tcp{Detect 1 bit in each iteration.}
\For{\text{$i=0;$ $i < n;$ $i++$}}{
\syncc{};\\
Access line0 and time the access;\\
\eIf{$access\_time > Th0$} {
        Received a bit ``1'';
    }
    {
        Received a bit ``0'';
    }
}
\caption{Prefetch+Load Covert Channel}
\label{alg1}
\end{algorithm}

\begin{algorithm}[t]
\scriptsize
\vspace{0.02in}
\SetKwFunction{sync}{sync\_with\_receiver}
\SetKwFunction{syncc}{sync\_with\_sender}
\textbf{line0}: the shared cache line between the sender and receiver\\
\textbf{message[n]}: the n-bit long message to transfer on the channel\\
\textbf{Th0}: the timing threshold on \texttt{PREFETCHW} to distinguish M and S states \\
\vspace{-0.02in}
--------------------------------------------------------------------------------------\\
\vspace{-0.02in}
\textbf{\scriptsize Sender Algorithm}\\
\vspace{-0.03in}
--------------------------------------------------------------------------------------\\
\tcp{Send 1 bit in each iteration.}
\For{\text{$i=0;$ $i < n;$ $i++$}}{
\sync{};\\
\eIf{$message[i] == 1$} {
        Load line0;
    }
    {
        Do not load;
    }
}

--------------------------------------------------------------------------------------\\
\vspace{-0.02in}
\textbf{\scriptsize Receiver Algorithm}\\
\vspace{-0.03in}
--------------------------------------------------------------------------------------\\
\tcp{Detect 1 bit in each iteration.}
\For{\text{$i=0;$ $i < n;$ $i++$}}{
\syncc{};\\
Prefetch line0 and time the prefetch;\\
\eIf{$prefetch\_time > Th0$} {
        Received a bit ``1'';
    }
    {
        Received a bit ``0'';
    }
}

\caption{Prefetch+Prefetch Covert Channel}
\label{alg2}
\end{algorithm}

\subsection{Prefetch+Prefetch Attack}

Our second attack, Prefetch+Prefetch, is based on Observation 2. As shown in Algorithm~\ref{alg2}, in each iteration of the attack, the sender transmits ``1'' by loading the shared cache line, or transmits ``0'' by idling. After this, the receiver performs \texttt{PREFETCHW} on the shared cache line and times the prefetch to decode the bit: when the sender sends ``1'', it takes longer for the receiver to prefetch than when the sender sends ``0''. 
Prefetch+Prefetch follows the same synchronization method with Prefetch+Load.


    
    



\section{Prefetch-Based Side Channel Attacks}
\label{side_channel}
\subsection{Basic Idea and Assumptions}
In the Prefetch+Prefetch covert channel attack, the sender is sending the signal by ``accessing (or not) the shared data''. Thus, this attack can be directly applied as a side channel attack to leak a victim's \textit{access pattern} on the shared data: the victim is the sender, and the attacker is the receiver. This leakage (the victim's access pattern) is same as the one in previous cache attacks (e.g.,~\cite{ff, fr, coherence_attack, template}). 

However, Prefetch+Load cannot be directly used as a side channel, because the sender is transmitting the signal by ``prefetching (or not) the shared data''. In other words, the attacker (receiver) can only detect the victim's (sender's) prefetch patterns on the shared data. Since software prefetch is not as common as memory accesses in real-world applications, the attack opportunities are limited. However, we can modify the attack slightly to make it work more generally.

We term the new attack \textit{Prefetch+\textbf{Re}load}. The attacker prefetches the shared data to pre-set the coherence state, and then waits for the victim  to possibly access this data. Later the attacker reloads the data (using a different thread on a different core, explained later) and uses the timing information to learn the current coherence state of the data, which leaks whether the victim has loaded this data (thus changing the coherence state). Different than Prefetch+Load, in Prefetch+Reload, the attacker needs to have two threads running on different cores.

\noindent
\textbf{Threat model.}
We assume a similar threat model as the one for the covert channels. First, the attacker is an unprivileged process that can 1) run on the same processor with the victim and 2) share data with the victim (e.g., through a shared library). The attacker aims at leaking the victim's access pattern on a shared data block, as in~\cite{fr, ff}. In addition, the attacker can launch his thread(s) on different core(s) than the victim's. 

For Prefetch+Reload, the attacker needs to have two threads running on different physical cores; but for Prefetch+Prefetch, there is still only one thread required in the attacker's process, which is the same setup as the covert channel attacks.

\subsection{Prefetch+\textbf{Re}load Attack}

\begin{figure}[!t] \centering
 \begin{minipage}[b]{0.99\columnwidth}
\centering
\includegraphics[width=0.99\columnwidth]{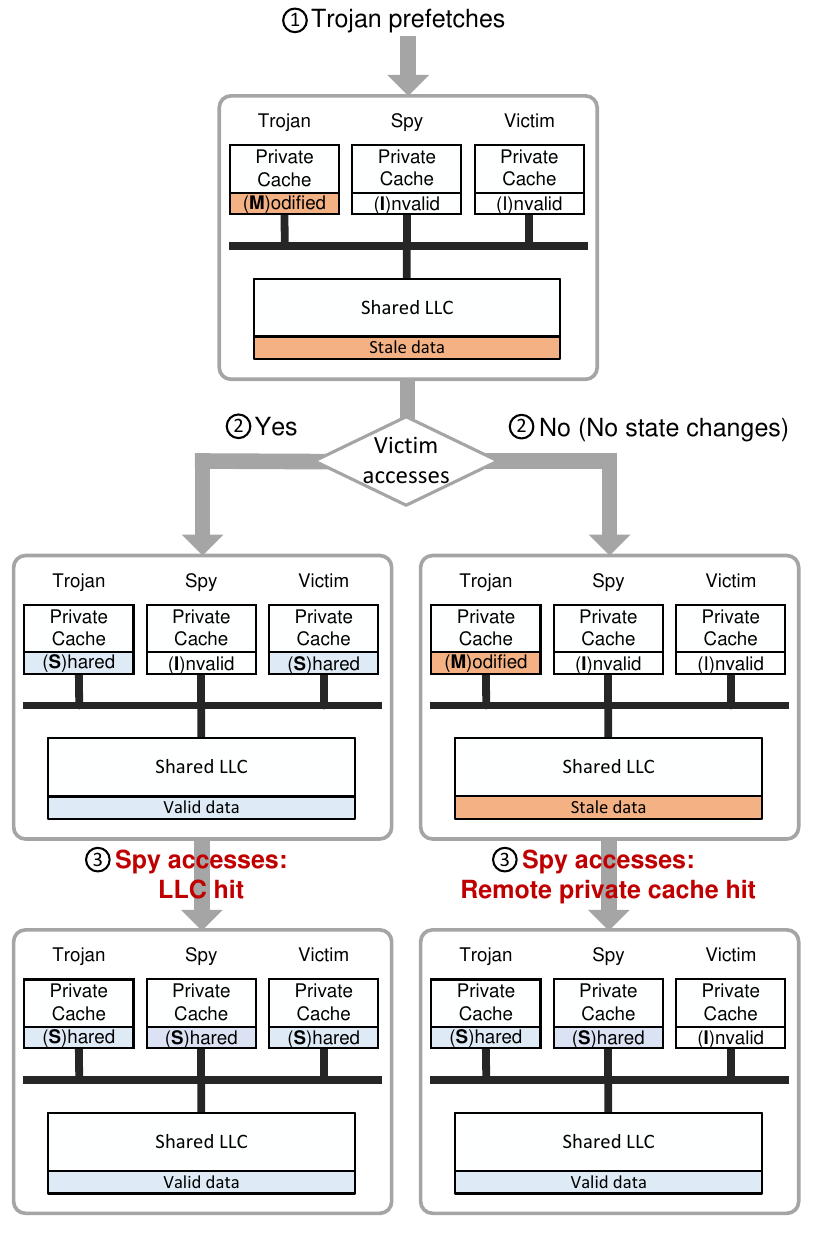}
\end{minipage} 
\caption{The details of the three steps in Prefetch+Reload. } \label{pr} 
\end{figure}

In this attack, we assume that the attacker controls two threads named \textit{Trojan} and \textit{Spy}. Trojan and Spy should be located on different cores, which are also both different than the victim's core, i.e., Trojan, Spy, and the victim all run on different cores. As mentioned in Section~\ref{cache}, the execution times of a remote private cache hit and an LLC hit are different. The Prefetch+Reload attacker uses this timing difference to observe cache state changes caused by the victim's accesses. Specifically, before the victim accesses the target shared cache line, Trojan executes \texttt{PREFETCHW} on this cache line, which invalidates the copies of this cache line in the victim's and Spy's private caches (if they exist), and places a copy of this cache line  (in M state) in Trojan's private cache, as shown in Step 1 of Figure~\ref{pr}. Then, if the victim accesses this cache line, according to MESI, the coherence state changes from M to S, and the copy of this cache line in the LLC is updated (although the content did not change, see Section~\ref{cache}) and is now valid (Step 2 in the left path of Figure~\ref{pr}).

Unfortunately, Trojan cannot observe this state change caused by the victim's access: if Trojan accesses (reloads) this cache line, he will get a private cache hit, 
no matter if the victim accessed this line or not. This is because the victim's read does not invalidate 
the copy in Trojan's private cache (Step 2 in the left path of Figure~\ref{pr}).
However, Spy is able to distinguish whether the victim accessed this cache line. Trojan's original \texttt{PREFETCHW} invalidated the copy in Spy's private cache. Thus, if Spy now accesses this cache line, he will get an LLC hit if the victim has accessed this cache line after Trojan's prefetch (Step 3 in the left path of Figure~\ref{pr}); otherwise, he will get a remote private cache hit (Step 3 in the right path of Figure~\ref{pr}). We recall that Spy can distinguish these two situations by timing the access (the difference is over 30 cycles on our desktop processor).
Based on this, we build Prefetch+Reload. Similar to previous cache attacks, each iteration in this attack contains three steps, as shown in Figure~\ref{pr}:
\begin{steps}[leftmargin=*]
    \item Trojan performs \texttt{PREFETCHW} on the target cache line and becomes the exclusive owner of this cache line.
    \item The attacker waits for the victim's behavior: if the victim accesses this cache line, its coherence state will become S, meaning the copy in the LLC is now valid.
    \item Spy accesses this cache line and times the access to determine it was a remote private cache hit or an LLC hit. If it was a remote private cache hit, then the victim did not access this cache line; otherwise the victim did access.
\end{steps}


\noindent
\textbf{LLC presence.} Prefetch+Reload requires that the target shared cache line is present in the LLC, so that Spy can get an LLC hit in Step 3, if the victim has accessed this cache line. This is naturally true for inclusive LLCs, since all the cache lines in the private cache are also present in the LLC. However, it is not guaranteed for non-inclusive LLCs. In those caches, a cache line is directly brought into the private cache when loaded from DRAM, bypassing the LLC; it usually goes to the LLC when evicted from the private cache due to cache replacement~\cite{skylake-sp,non-inclusive-attack}. 
Thus, strictly speaking, it is the attacker's responsibility to ensure the presence of this cache line in the LLC, if it is non-inclusive. For example, before the attacker starts the attack loop, he can first build set conflicts in his private cache to evict this cache line to the LLC. 

In fact, empirically we found that in Step 1, when \texttt{PREFETCHW} invalidates the copies of the target cache line in Spy and the victim's private caches, this cache line will be placed in the LLC if it does not already exist. Therefore, in practice the attacker does not need to explicitly place this cache line in the LLC.

 \begin{figure*}[!t] \centering
 \begin{minipage}[b]{\textwidth}
\centering
\includegraphics[width=\columnwidth]{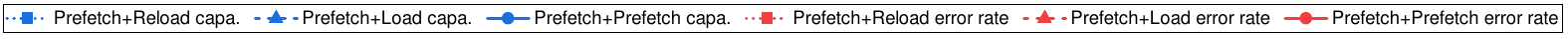}
\end{minipage} 
 \vspace{-0.3in}
\end{figure*}

\begin{figure*}[!tb]
\centering
  \subfigure[Intel Core i7-6700]{
    \begin{minipage}[b]{0.238\textwidth}
    \centering
      \includegraphics[width=\columnwidth]{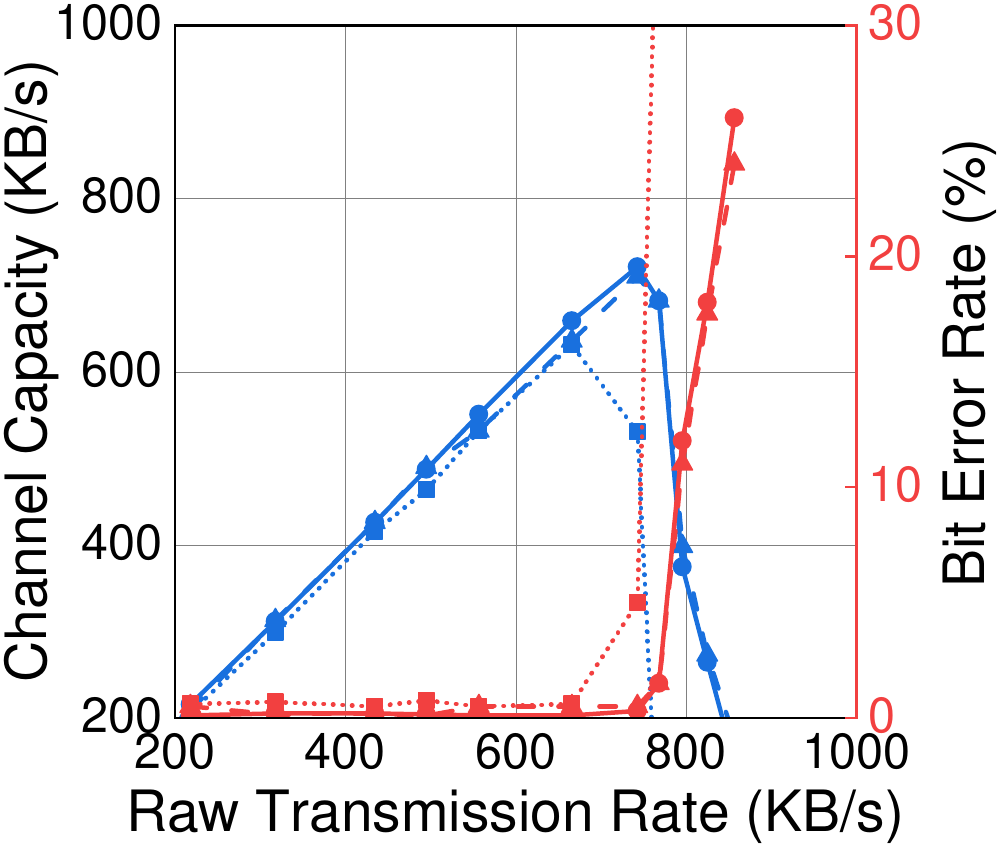}
    \end{minipage}}
      \subfigure[Intel Core i7-7700K]{
    \begin{minipage}[b]{0.238\textwidth}
    \centering
      \includegraphics[width=\columnwidth]{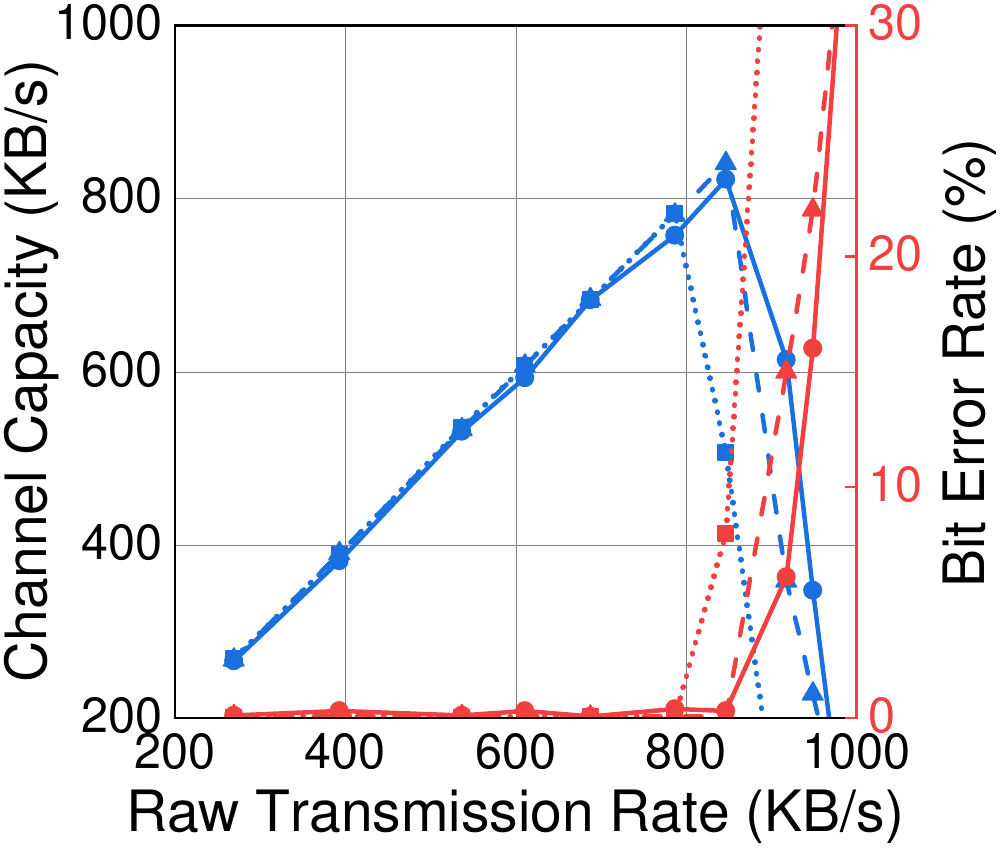}
    \end{minipage}}
      \subfigure[Intel Xeon Platinum 8124M]{
    \begin{minipage}[b]{0.238\textwidth}
    \centering
      \includegraphics[width=\columnwidth]{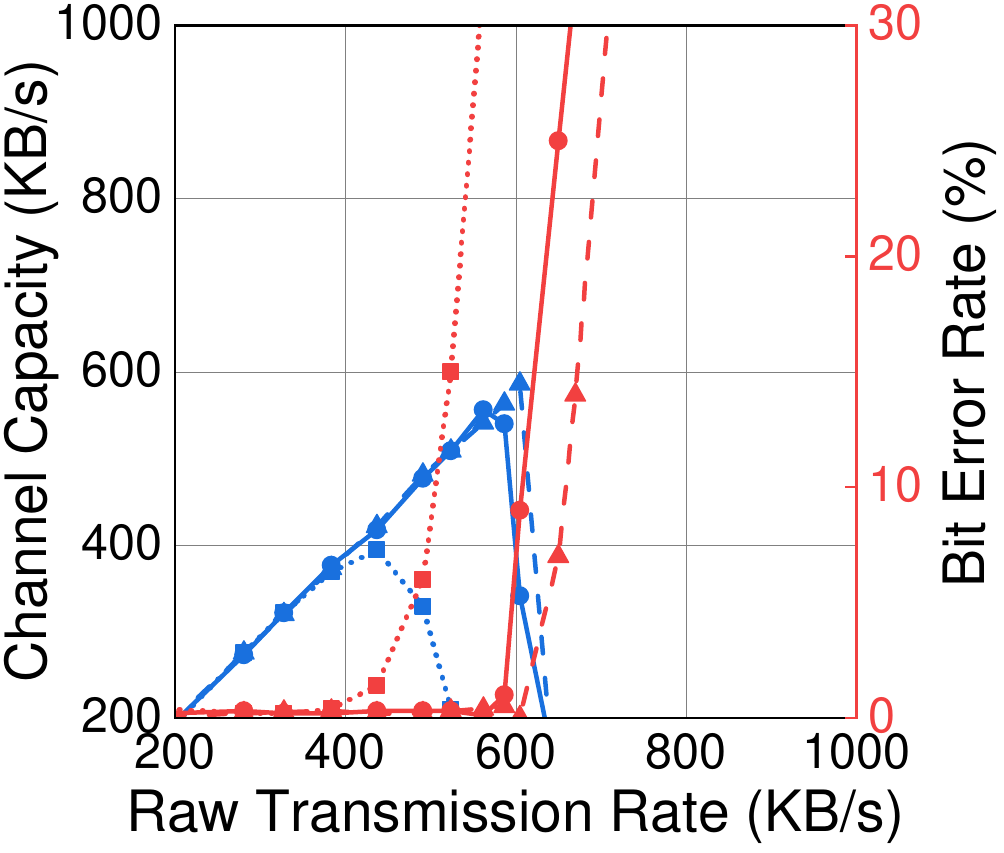}
    \end{minipage}}
      \subfigure[Intel Xeon Platinum 8151]{
    \begin{minipage}[b]{0.238\textwidth}
    \centering
      \includegraphics[width=\columnwidth]{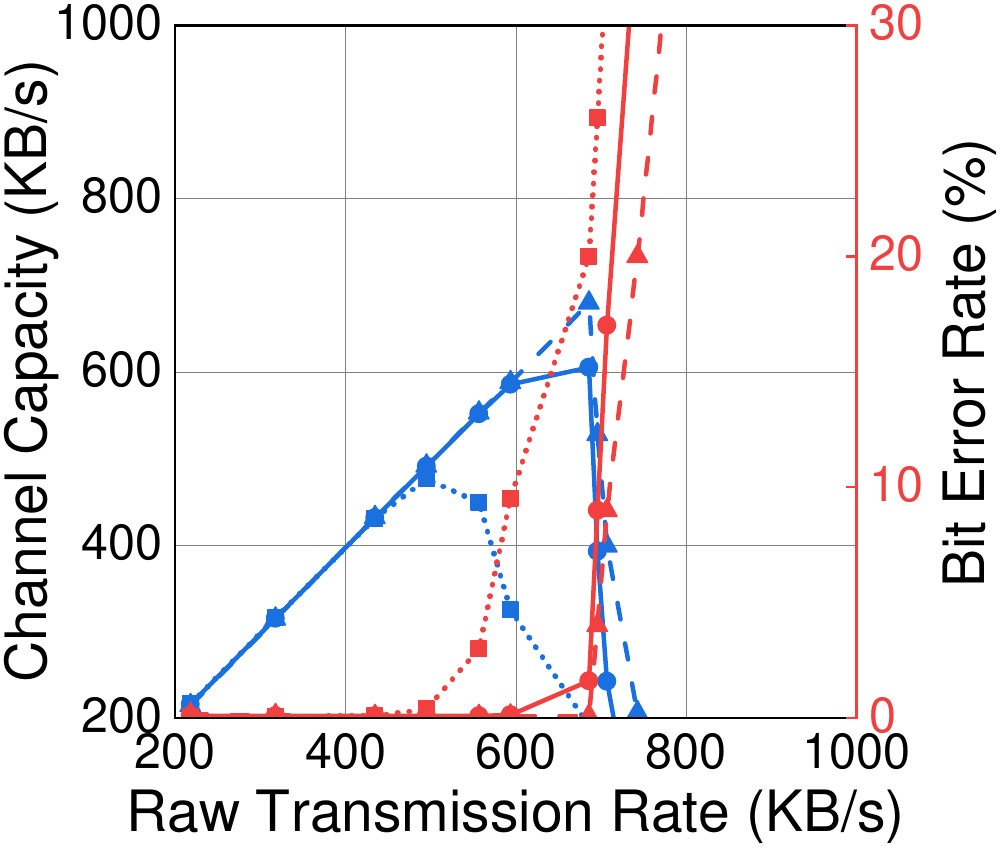}
    \end{minipage}}
    \caption{The capacities and bit-error-rates of the prefetch-based channels on various Intel processors.} 

\label{covert_result1} 
\end{figure*}

\subsection{Prefetch+Prefetch Attack}
Following the Prefetch+Prefetch covert channel attack, we also build the Prefetch+Prefetch side channel attack. 
The attacker learns if the victim accessed the shared cache line by timing \texttt{PREFETCHW}. 
In contrast to the Prefetch+Reload side channel attack, each iteration in this attack only has two steps:

\begin{steps}[leftmargin=*]
    \item The attacker prefetches the target shared cache line using \texttt{PREFETCHW}, and times this operation to learn whether the victim accessed this cache line in the last iteration. 
    \item The attacker waits for the victim's behavior.
\end{steps}

As explained earlier in Section~\ref{characterization}, in Step 1 above, if the victim accessed this cache line,  \texttt{PREFETCHW} executes slower; if the victim did not access, \texttt{PREFETCHW} executes faster.

In contrast to most previous cross-core cache attacks, which can only work on the LLC and require repeatedly evicting the target cache line to DRAM (e.g., Flush+Reload), the proposed prefetch-based attacks work on the private cache. Thus, the target cache line is always kept in the on-chip cache hierarchy. Compared to cross-core LLC attacks, cross-core private cache attacks have two benefits. First, higher bandwidth, since cache accesses are fast and are usually much faster than DRAM accesses. This is especially important when the attacks are used as covert channels. Second, stealthier, since there are less cache misses, especially LLC misses involved in the attacks~\cite{detect}. 
\textit{To the best of our knowledge, the proposed prefetch-based attacks are the first cross-core private cache side channel attacks that can work regardless of the LLC inclusivity.}

\section{Evaluation}
\label{experiment}

We evaluate the proposed covert channel and side channel attacks on modern Intel processors. 
For covert channel attacks, we evaluate the channel capacities, comparing them to previous cache covert channels on CPU.  
For side channel attacks, we demonstrate how they can be used to leak information from common applications.  
In addition, we also show how our attacks strengthen transient execution attacks.

\subsection{Evaluation of Prefetch-Based Covert Channel Attacks}
We implement Prefetch+Load, Prefetch+Prefetch, and Prefetch+Reload on four Intel processors, including two desktop processors and two server processors. Note that although Prefetch+Reload is introduced as a side channel attack in Section~\ref{side_channel}, it can be a covert channel attack as well.  Table~\ref{processor} lists the specifications of the four tested processors. 
The two desktop processors have inclusive LLCs, and the server processors have non-inclusive LLCs.

We use one shared cache line between the sender and receiver to transmit secrets. Although using more shared cache lines or using channel coding techniques (e.g.,~\cite{ff}) may further improve the channel capacity~\cite{ff, coherence_attack}; here we do not include them since we aim at showing the conservative results (i.e., the lower bounds).

We measure the channel capacity and bit error rate of each attack, under different transmission intervals. Although the raw transmission rate increases when decreasing the transmission interval, the bit error rate may also increase, especially when the interval is too short. To find the best transmission rate, we use the channel capacity metric (as in~\cite{drama, ring}). This metric is computed by multiplying the raw transmission rate with $1-H(e)$, where $e$ is the bit error rate and $H$ is the binary entropy function.
The results are shown in Figure~\ref{covert_result1}. The bit error rates of all three attacks stay low (lower than 0.6\%) and are almost constant, when the raw transmission rate is under a threshold (e.g., 660 KB/s for Prefetch+Reload in Figure~\ref{covert_result1}(a)). Thus, the channel capacity increases proportionally to the raw transmission rate. It reaches the peak when the raw transmission rate is around this threshold. Beyond this threshold, the increasing error rate causes a decrease in the channel capacity. The peak channel capacities of the three attacks are summarized in Table~\ref{covert_maximum_rate}. 
Prefetch+Reload always has lower capacity than the other two attacks because more cache operations are involved in each iteration of Prefetch+Reload.

\begin{table}[!h]
\vspace{0.08in}
\begin{center} \caption{The specifications of the tested processors.} \label{processor} 
\setlength\arrayrulewidth{1.0pt}
\renewcommand{\arraystretch}{1.2}
    \setlength{\tabcolsep}{1.0mm}{ \begin{tabular}{lrrrr} 
    \hline
     &\multicolumn{2}{l}{Desktop processors} & \multicolumn{2}{l}{Server processors} \\
     \cline{2-5}
     Platform & \vtop{\hbox{\strut Core}\hbox{\strut i7-6700}}& \vtop{\hbox{\strut Core}\hbox{\strut i7-7700K}}& \vtop{\hbox{\strut Xeon Platinum}\hbox{\strut 8124M}} & \vtop{\hbox{\strut Xeon Platinum}\hbox{\strut 8151}} \\ \hline
     Microarchitecture & Skylake &Kaby Lake  & Skylake-SP & Skylake-SP\\ 
     Num of cores & 4 & 4 & N/A\footnotemark & N/A \\ 
     Frequency & 3.4 GHz & 4.2 GHz & 3.0 GHz & 3.4 GHz \\ 
     LLC type & Inclusive & Inclusive & Non-inclusive & Non-inclusive\\ 
     \hline
\end{tabular}} \end{center}  
\end{table}

\footnotetext{We use Intel Xeon Scalable processors on Amazon AWS EC2 platform, and we leased four physical cores on the tested processors for our experiments.}


\textit{Our prefetch-based attacks are faster than almost all existing cache attacks on x86 CPUs.} 
First, for attacks tested on desktop processors, the ring interconnect contention based attack~\cite{ring} is reported with a very high capacity which is 518 KB/s on a 4.0 GHz desktop processor. Flush+Reload and Flush+Flush have capacities of 298 KB/s and 496 KB/s on a 3.6 GHz desktop processor~\cite{ff}, respectively.
Prime+Scope~\cite{ps}, the optimized attack for Prime+Probe, achieves 438 KB/s on a 3.5 GHz desktop processor. Second, most of the attacks that were tested on server processors, including the L1 LRU attack~\cite{lru1}, the directory Prime+Probe attack~\cite{non-inclusive-attack}, and the Flush+Coherence attack~\cite{coherence_attack} have capacities of less than 200 KB/s. The directory version of Prime+Scope achieves 387 KB/s.

\begin{table}[!t]
\vspace{0.06in}
\begin{center} \caption{The maximum capacities of the prefetch-based channels.} \label{covert_maximum_rate} 
\setlength\arrayrulewidth{1.0pt}
\renewcommand{\arraystretch}{1.4}
    \setlength{\tabcolsep}{1.0mm}{ \begin{tabular}{lllll} 
    \hline
     & \multicolumn{2}{l}{Desktop processors} & \multicolumn{2}{l}{Server processors} \\
     \cline{2-5}
     Platform & \vtop{\hbox{\strut Core}\hbox{\strut i7-6700}\hbox{\strut (3.4 GHz)}}& \vtop{\hbox{\strut Core}\hbox{\strut i7-7700K}\hbox{\strut (4.2 GHz)}}& \vtop{\hbox{\strut Xeon Platinum}\hbox{\strut 8124M}\hbox{\strut (3.0 GHz)}} & \vtop{\hbox{\strut Xeon Platinum}\hbox{\strut 8151}\hbox{\strut (3.4 GHz)}} \\ \hline
     
     Prefetch+Reload & 631 KB/s & \textbf{782 KB/s}  & 394 KB/s & 476 KB/s\\ 
     Prefetch+Load & 709 KB/s & \textbf{840 KB/s} & 586 KB/s & 680 KB/s \\ 
     Prefetch+Prefetch & 721 KB/s & \textbf{822 KB/s} & 556 KB/s & 605 KB/s \\ 
     \hline
\end{tabular}} \end{center}  
\end{table}

To the best of our knowledge, our attacks are only slower than Streamline~\cite{streamline}. This attack claims to achieve a capacity of 1801 KB/s. However, it has such a high channel capacity because the sender and receiver use 64 MB shared data to transmit secrets; our results are based on one shared cache line (64 B).




\subsection{Evaluation of Prefetch-Based Side Channel Attacks}

\subsubsection{\textbf{Side Channel Attack on Cryptographic Code}}
Our first attack targets cryptographic libraries, where the access patterns to some instructions are related to the value of the cryptographic key. 
More specifically, we target the square-and-multiply algorithm~\cite{square_algorithm} which is used in GnuPG 1.4.13 for ciphers such as RSA~\cite{rsa} and ElGamal~\cite{Elgamal}: leaking the exponent \textit{e} of this algorithm leaks the private key. As shown in Algorithm~\ref{rsa_alg}, in each loop iteration,  it first executes a \texttt{sqr} and a \texttt{mod} instruction. Then, if the exponent bit is ``1'', a \texttt{mul} and another \texttt{mod} instruction are executed; otherwise they are skipped. Thus, by monitoring the access pattern to the cache lines that contain \texttt{sqr} and \texttt{mul}, the attacker is able to leak each bit of the exponent \textit{e} and therefore the decryption key.

\begin{algorithm}[!h]
\scriptsize
\vspace{0.02in}
\SetKwFunction{sync}{sync\_func}
\textbf{Input}: base $b$, modulo $m$, exponent $e = (e_{n1} ... e_{0})_2$\\
\textbf{Output}: $b^{e}$ mod $m$\\
\vspace{0.06in}

$r \leftarrow 1$\\
\For{\text{$i=n-1;$ $i >= 0;$ $i--$}}{
$r \leftarrow r^2 \mod n$ \\
\If{$e_i == 1$} {
        $r \leftarrow r*b \mod n$
    }

}
\caption{Square-and-multiply Exponentiation}
\label{rsa_alg}
\end{algorithm}




\noindent
\textbf{Implementation.} As done in the Flush+Reload attack on GnuPG~\cite{fr}, we use \texttt{mmap} to map the pages that contain \texttt{sqr} and \texttt{mul} into the attacker's address space. Note that during the execution of the victim (GnuPG), the cache lines containing those instructions are brought into the victim's L1 instruction cache (L1I cache). However, since we map the instruction pages as data blocks in the attacker's address space, the same cache lines containing those instructions are brought to the attacker's L1D cache. Thus, although \texttt{PREFETCHW} can only prefetch cache lines into L1D cache, it can still leak the victim's access patterns to instructions. 

 \begin{figure}[!h] \centering
 \begin{minipage}[b]{0.99\columnwidth}
\centering
\includegraphics[width=\columnwidth]{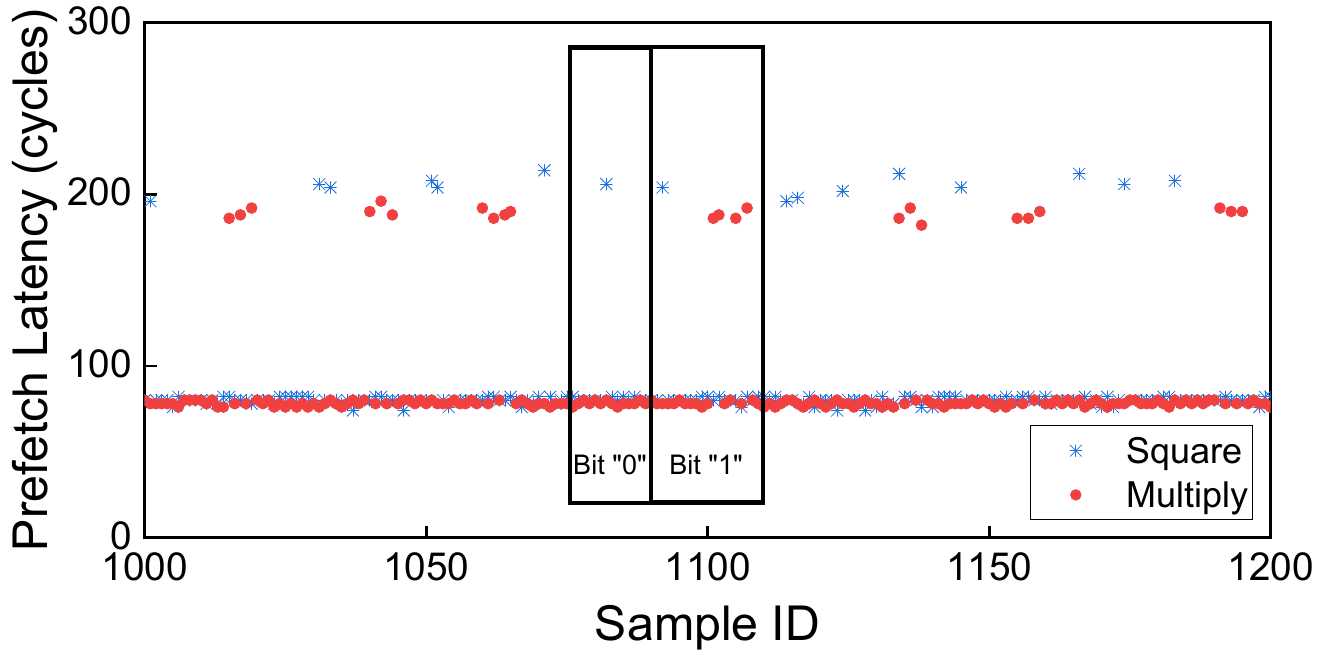}
\end{minipage} 
\caption{A segment of the prefetch latencies measured in  Prefetch+Prefetch while attacking GnuPG; part of the the exponent $e$ shown here is ``111001011001''. } \label{rsa_result} 
\end{figure}

\noindent
\textbf{Results.} For simplicity, we only show the attack results of Prefetch+Prefetch on the Intel Xeon Platinum 8151 processor. However, we have performed this attack on other processors listed in Table~\ref{processor} too, using both Prefetch+Prefetch and Prefetch+Reload. Here we use a waiting latency of 500 cycles in each iteration of Prefetch+Prefetch.
Figure~\ref{rsa_result} shows the timing measurement results from the attacker for 200 samples: a lower prefetch latency (less than 100 cycles) indicates that the victim did not access the target cache line during the last iteration; a higher prefetch latency (around 200 cycles) means the victim did access. 
As explained above, an access to \texttt{sqr} followed by an access to \texttt{mul} indicates a bit ``1'', and two consecutive accesses to \texttt{sqr} (one from the current iteration, one from the next iteration) indicate a bit ``0'' (in the current iteration). Thus, part of the exponent \textit{e} shown in Figure~\ref{rsa_result} is ``111001011001''. The average attack accuracy is 96.2\%.



\subsubsection{\textbf{Side Channel Attack on Keystroke Timing}}
Our second attack focuses on leaking the precise timing information of keystrokes, i.e., detecting when a keyboard input occurs. This leakage is important since it can assist reconstructing typed words from users~\cite{key1, key2, key3}. Previous work shows that certain functions in graphics libraries are called when a keystroke happens (e.g.,~\cite{ndss_key,ff}). Thus, we can monitor the accesses to the cache lines containing these functions to detect keystrokes.

\noindent
\textbf{Implementation.}
We attack an address in the shared GDK library which is invoked when processing keystrokes. 
Specifically, we launch gedit as the victim, and input keystrokes in it. At the same time, we run the prefetch-based attacks to monitor accesses to the address selected in the GDK library, and record the timing measurement results. The attacker process raises an alarm when a keystroke is detected.

\begin{figure}[!h] \centering
      \includegraphics[width=0.99\columnwidth]{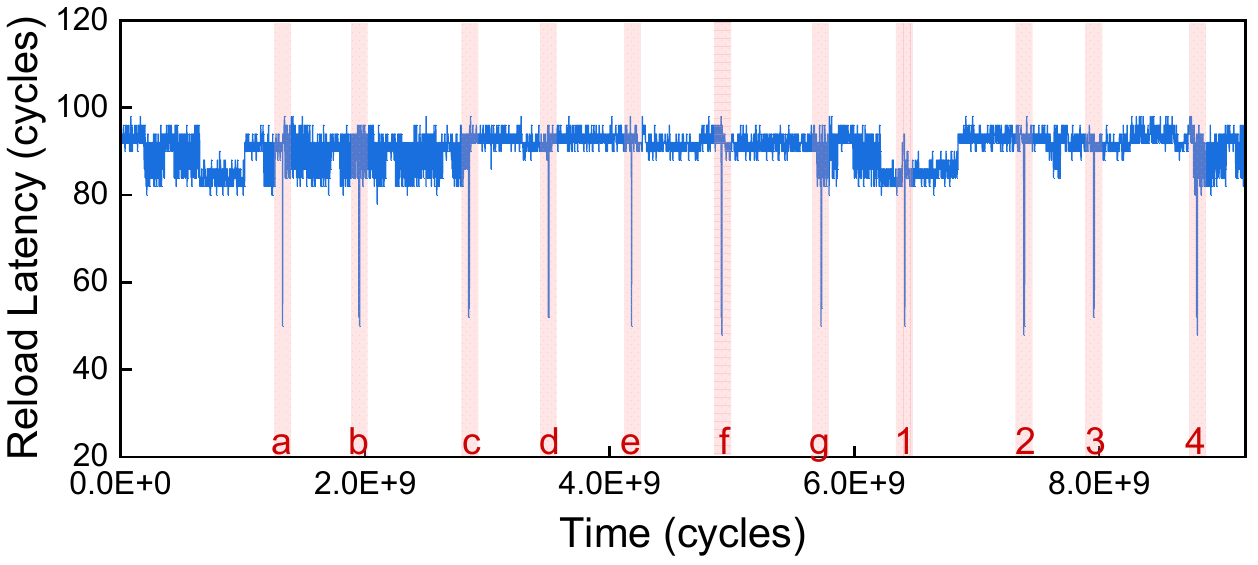}
\caption{The access latencies measured in Step 3 of Prefetch+Reload when a user types ``abcdefg1234'' in gedit; we monitor address 0x7b980 of libgdk.so.\protect\footnotemark}
\vspace{0.13in}
\label{keystroke_result}
\end{figure}

\noindent
\textbf{Results.}
Figure~\ref{keystroke_result} shows the timing trace collected by Prefetch+Reload when the user is typing ``abcdefg1234'' in gedit, on our Intel Core i7-6700 processor. Again, the attack has been done on the other desktop processor too (but not on the server processors since EC2 instances do not come with GUI). As one can observe, when a keystroke occurs, the reload operation (in Step 3 of Prefetch+Reload) takes around 50 cycles to finish; it takes over 80 cycles to reload when there is no keystroke. This significant timing difference makes keystrokes very detectable. During the attack, we observe zero false positives and zero false negatives. 


\footnotetext{We found the appropriate library and address to monitor following the method in prior work~\cite{template}. }

\subsubsection{\textbf{Windowless Prefetch+Prefetch}}

Using the terminology in prior work~\cite{ps}, \texttt{PREFETCHW} has two important properties: 1) \texttt{PREFETCHW} is preserving, meaning the measurement (prefetching and timing the prefetch) does not change the state in the absence of the victim's event; 2) \texttt{PREFETCHW} is also concurrent, meaning it detects the events that temporally overlap with it. 
With these two features, Prefetch+Prefetch can be used in a windowless way (no waiting window between two consecutive prefetches is necessary). We verify this using the following experiment.

\begin{figure}[!t] \centering
 \begin{minipage}[b]{0.93\columnwidth}
\centering
\includegraphics[width=\columnwidth]{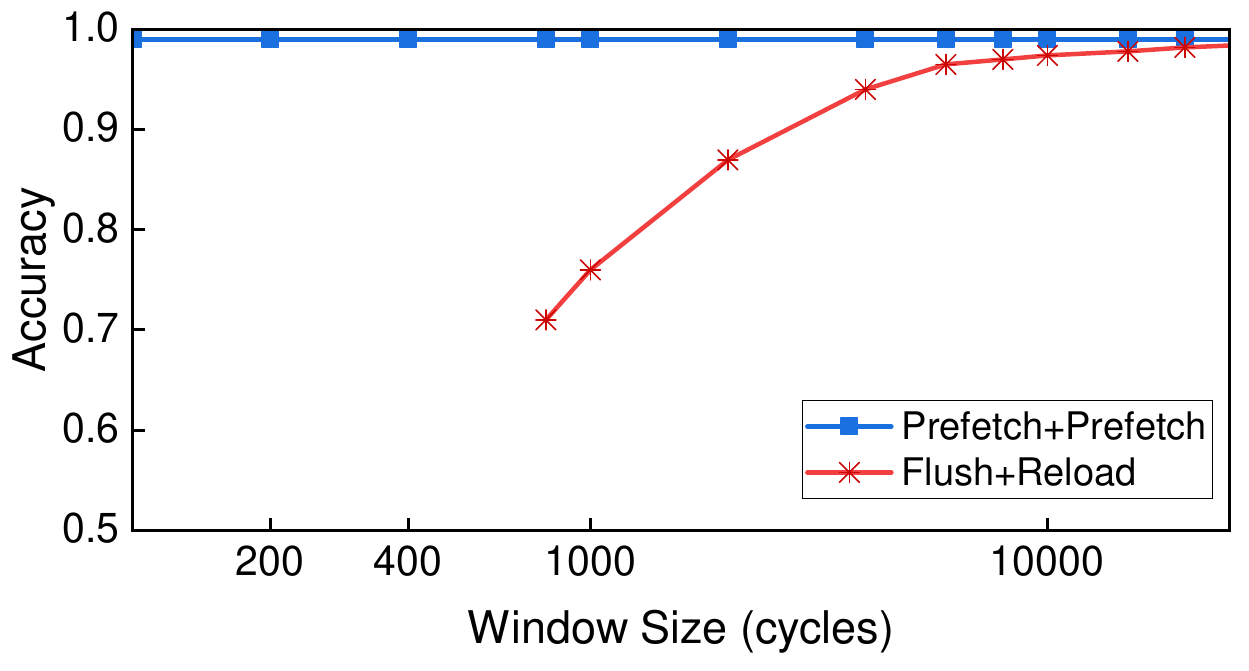}
\end{minipage} 
\caption{The accuracy of Prefetch+Prefetch and Flush+Reload on our Intel Core i7-6700 processor, with different waiting window sizes. } \label{resolution} 
\end{figure}

We use two processes, the victim and attacker. The victim process first waits a random amount of time, and then triggers an event (accessing the target shared cache line). This process terminates after triggering the event. The attacker process runs Prefetch+Prefetch with a waiting window in each attack iteration to detect the victim's event. The attacker process terminates after detecting the event or after the victim process terminates. We run this experiment with different window sizes and repeat the experiment for 1000 times for each window size. Figure~\ref{resolution} shows the attacker's detection accuracy on our Intel Core i7-6700 processor. Note that the results on other processors in Table~\ref{processor} are similar. For comparison, we also show the accuracy of Flush+Reload on the same processor. For Prefetch+Prefetch, the attacker's detection accuracy does not change when the window size varies; the attacker always has a very high detection accuracy which is around 1. This indicates that Prefetch+Prefetch, unlike prior attacks such as Flush+Reload, can always be used as a windowless attack. Such a windowless attack has much higher temporal resolution than a windowed attack since the latter's resolution is bounded by the window size. For example, to reach 95\% detection accuracy, Flush+Reload needs a waiting window with over 4000 cycles.

\subsection{Prefetch-Based Channels in Transient Execution Attacks}

Transient execution attacks such as Spectre~\cite{spectre} and Meltdown~\cite{meltdown} usually require a microarchitectural covert channel to transfer the secrets to the attacker. Currently, most transient execution attacks (e.g.,~\cite{spectre, meltdown, ridl, fallout, zombie}) use the Flush+Reload channel because it is simple, reliable, and ubiquitous. 
Here we demonstrate that prefetch-based channels can also work with transient execution attacks to leak secrets, and may even work better than Flush+Reload. We use Spectre v1 as an example to show the details and benefits of using prefetch-based channels in transient execution attacks.

\noindent
\textbf{Higher bandwidth.}
When using Flush+Reload, the sender operation in Spectre is a memory access where the address depends on the secret value.
Since Prefetch+Reload and Prefetch+Prefetch use the same sender function as Flush+Reload, a victim program vulnerable to Spectre with Flush+Reload is also vulnerable to Spectre with Prefetch+Reload and Prefetch+Prefetch. We have verified this using the Spectre v1 PoC code~\cite{spectre_poc}. We modify it accordingly such that Prefetch+Reload or Prefetch+Prefetch is used in the attacker code; the victim remains the same. In addition, as observed in prior work~\cite{meltdown}, the leakage rate of a transient execution attack is significantly affected by the capacity of the covert channel used in the attack. Since Prefetch+Reload and Prefetch+Prefetch have much higher capacities than Flush+Reload, Spectre works faster with these two channels. For example, on our Intel Core i7-6700 processor, when leaking an 8-bit secret in each transmission, the leakage rate of Spectre is 3.02 times and 1.61 times as fast when using Prefetch+Prefetch and Prefetch+Reload, respectively, as compared to Flush+Reload. The results on other processors are shown in Appendix A.

\noindent
\textbf{More leakage in the transient window.}
When using Spectre with Flush+Reload, the data access for sending (encoding) the secret in the transient window is a \textit{slow DRAM access}, since this data was flushed by the attacker. In contrast, the data access for secret encoding is a \textit{remote private cache hit} when using Prefetch+Reload or Prefetch+Prefetch, which is usually faster than a DRAM access. 
This indicates that within the same transient window, more encoding operations can be performed using the two prefetch-based channels than Flush+Reload, and thus more secrets may be leaked.
An example Spectre v1 gadget that can benefit from this is shown in Listing~\ref{list_spectre}. There are n operations in the branch, where each operation accesses a secret and encodes it to a cache index. The secrets are array1[x] to array1[x+n] (when x is out of bounds); each of the secrets is encoded to an index of a sub-array in array2. The more of these n operations we can perform in the transient window, the more secrets we can leak out at once.

\begin{figure}[h]
\centering
\begin{minipage}[!b]{0.97\columnwidth}
\scriptsize
\begin{lstlisting}[style=bb]
        if (x+n < array1_size)
        {
            y0 = array2[0][array1[x] * 4096];
            y2 = array2[1][array1[x+1] * 4096];
            ...
            yn = array2[2][array1[x+n] * 4096];
        }

\end{lstlisting}
\captionof{lstlisting}{The Spectre v1 code example when a bounds check is followed by multiple secret accessing and encoding operations. This code is essentially a for loop in a conditional branch; we show the unrolled version for clarity.}
\label{list_spectre}
\end{minipage}
\end{figure}

\begin{figure}[!h] 
\vspace{0.1in}
    \centering
  \subfigure[Intel Core i7-6700]{
    \begin{minipage}[b]{0.47\columnwidth}
      \includegraphics[width=0.99\columnwidth]{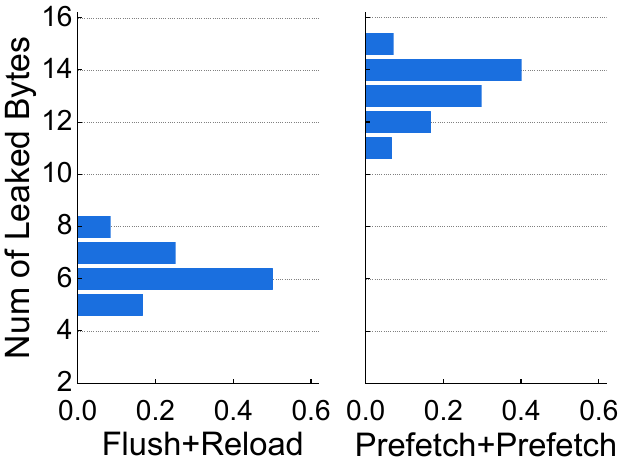}
    \end{minipage}}
  \subfigure[Intel Core i7-7700K]{    
    \begin{minipage}[b]{0.47\columnwidth}
      \includegraphics[width=0.99\columnwidth]{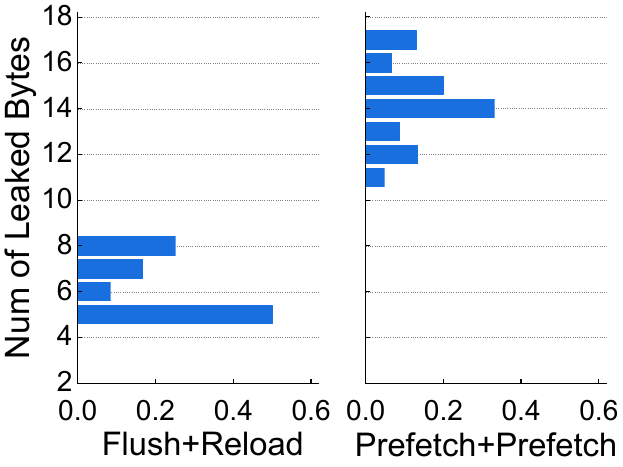}
    \end{minipage}}  
  \subfigure[Intel Xeon Platinum 8124M]{
    \begin{minipage}[b]{0.47\columnwidth}
      \includegraphics[width=0.99\columnwidth]{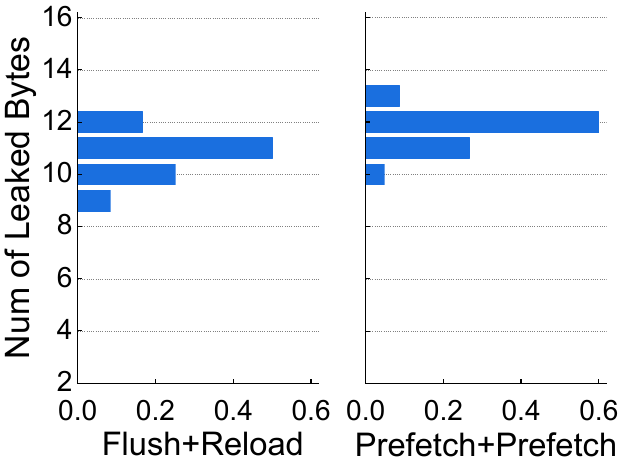}
    \end{minipage}}
  \subfigure[Intel Xeon Platinum 8151]{    
    \begin{minipage}[b]{0.47\columnwidth}
      \includegraphics[width=0.99\columnwidth]{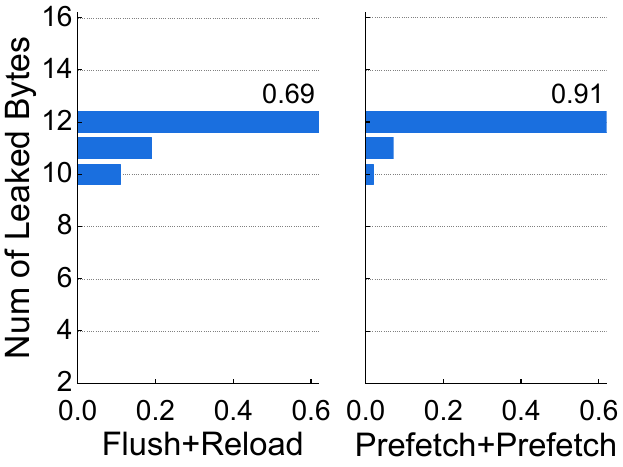}
    \end{minipage}}  
    \caption{The distributions of the amount of secret bytes that can be accessed and encoded in a transient window, when leaking by Flush+Reload and Prefetch+Prefetch, respectively.}%
    \label{spectre_result}%
  \vspace{0.1in}
\end{figure}

This gadget might be found in a victim; it is essentially the original Spectre v1 gadget with multiple secrets accessed and encoded in the branch (instead of one). 
Additionally, in the scenario where the attacker has control over the gadget (e.g., spectre-type-meltdown),\footnote{Spectre can be used for exception suppression in Meltdown.} the attacker can build such a gadget to leak multiple secrets in one transient window and thus accelerate the attack. 
We still prove this with the Spectre v1 PoC code and modify the attacker code to use Prefetch+Reload or Prefetch+Prefetch. We also modify the victim code to simulate the gadget in Listing~\ref{list_spectre} where $n$ secrets are accessed and encoded in the branch. We run this code and collect the amount of these $n$ secrets the victim can transmit within one transient window, and draw the histograms in Figure~\ref{spectre_result}. We omit the results
when leaking by Prefetch+Reload since its encoding stage is same as the one of Prefetch+Prefetch.

On the desktop processors, the victim can transmit up to 17 8-bit secrets speculatively when using Prefetch+Reload or Prefetch+Prefetch, while the victim can transmit at most 8 secrets when using Flush+Reload. However, on server processors, the amount of transmitted secrets when using prefetch-based channels is only slightly larger than the one when using Flush+Reload. This is because on these processors, the latency of a remote private cache hit is much longer, compared to desktop processors (160 cycles vs. 90 cycles). 
Note that although same-core private cache attacks, such as the L1 LRU attack~\cite{lru1}, can also achieve more secret encodings in a transient window than Flush+Reload, these attacks are less practical, because they are limited by the number of private cache sets. In these attacks, secret values are encoded into the cache set index instead of cache line index.

\noindent
\textbf{Other transient execution attacks.}
All of the three prefetch-based channels can be used in transient execution attacks when the attacker has full control of the gadget (e.g., Meltdown). As shown above, Prefetch+Reload and Prefetch+Prefetch has faster encoding operations than Flush+Reload, enabling more leakage in a transient window. The same is true for Prefetch+Load, since a remote private cache hit for \texttt{PREFETCHW} is usually faster than a DRAM access. In a Meltdown PoC with the three prefetch-based channels, we can reliably leak 8 bytes in the transient window on Our Intel Core i7-6700 processor; we can only leak 6.1 bytes on average when using Flush+Reload. An example gadget to achieve is shown in Appendix B.

\section{Discussion}
\label{discussion}

\subsection{Attack Reliability}
According to Intel~\cite{intel_optimization_manual}, a prefetch instruction will not fetch any data when the request buffer between the L1 and L2 cache is full. 
This may reduce the performance of the prefetch-based attacks, when SMT is available and a memory-intensive thread is located on the same core as the attacker thread. 
We verified this by running \texttt{stress -m 1} in a co-located thread (i.e., the hyperthread sibling) of the attacker thread: this causes many prefetch instructions from the attacker to be ignored, which significantly reduces the attack performance. For example, on our Intel Core i7-6700 processor, the channel capacity of Prefetch+Prefetch is reduced to 56 KB/s. However, SMT enables many security vulnerabilities (e.g.,~\cite{foreshadow}) and thus is often suggested to be disabled. In fact, if SMT is available, the attacker can always launch same-core private cache attacks instead. Our cross-core private cache attacks target the scenarios where same-core attacks are impractical or impossible.


\subsection{Prefetch-Based Attacks on AMD}
Modern AMD processors also support \texttt{PREFETCHW}; this instruction was  originally invented by AMD~\cite{3dnow}, and was later adopted by Intel. We performed the same experiments as the ones in Section~\ref{characterization} on AMD desktop and server processors. However, from our experiments, \texttt{PREFETCHW} does not cause any coherence state changes on data with read-only permission; it only works on data with write permission. Thus, we believe that AMD processors actually have permission checks for \texttt{PREFETCHW}.  

\subsection{Related Work}
\noindent
\textbf{Prefetch-based attacks.}
Gruss et al.~\cite{prefetch_attack} made two observations about prefetch instructions on Intel processors. They found that the execution time of a prefetch instruction, such as \texttt{PREFETCHT0}, leaks the translation levels of inaccessible kernel addresses. Using this, they built an attack to break Kernel Address Space Layout Randomization (KASLR).
They also observed that prefetch instructions change the cache state of inaccessible kernel memory, but recent work~\cite{prefetch_wrong} proved this incorrect. In fact, their observation is the result of transient execution caused by a Spectre gadget in the kernel, not the prefetch instruction.

Very recently, Lipp et al.~\cite{prefetch_attack_amd} observed that on AMD processors, the timing (and power consumption) of a prefetch instruction on an inaccessible kernel address leaks the translation level and TLB state of this address. They used this to break KASLR and leak kernel memory (with Spectre) on AMD processors.
These two prefetch attacks are orthogonal to our attacks. They focus on specifically attacking the kernel; we instead focus on building general cache timing attacks.   

Regarding attacks based on hardware prefetchers, Shin et al.~\cite{pre_security} 
attacked OpenSSL, leaking the private key by leveraging the Intel stride prefetcher. 
Rohan et al.~\cite{prefetch_hardware_attack2} reverse-engineered the stream
prefetcher on Intel processors, using it to build a covert channel.

\noindent
\textbf{Cache coherence vulnerabilities.}
Although we are the first to propose cross-core private cache side channel attacks leveraging cache coherence protocol invalidations, cache
coherence protocols have been exploited in many different attacks. 
Trippel et al.~\cite{spectreprime} discovered that a transient write may change the coherence state of the target data, which can be used as a covert channel in transient execution attacks.
In addition, previous studies~\cite{coherence_project_zero, spectre, coherence_rowhammer} 
mention that ``bouncing" cache lines between private caches may be used as a replacement for \texttt{CLFLUSH} or set conflicts in Spectre and Rowhammer attacks. However, in this method, coherence states are manipulated by write operations. This means it requires that at least part of the target cache line happens to contain writable data (unless Meltdown-RW~\cite{meltdown_rw,sysmatic} can be exploited). Unfortunately, as discussed in~\cite{coherence_rowhammer}, this requirement is impractical for general side channel attacks.  

Prior work~\cite{coherence_attack_amd,coherence_attack_arm}  built  cross-core attacks on AMD and ARM processors, respectively, based on cache coherence. An Evict+Reload attack on Intel processors with non-inclusive LLCs was proposed in~\cite{non-inclusive-attack}. In these three attacks, the attacker learns the victim's behavior by distinguishing between remote private cache hits and DRAM accesses. 
A variant of Flush+Reload attack on x86 processors was proposed in~\cite{coherence_attack}. It works by distinguishing between remote private cache hits and LLC hits. These  attacks are more general than the ones discussed earlier, but they all suffer from low bandwidth as DRAM accesses are involved in the attacks.





\subsection{Mitigations}

Our attacks can be prevented through modifications on the microarchitecture behavior of \texttt{PREFETCHW}. The complete protection is two-fold. First, \texttt{PREFETCHW} should perform write permission checks, just as a regular memory write instruction, and trigger a fault or ignore this instruction if the target data is not writable. Second, \texttt{PREFETCHW} should execute in constant time. These modifications may introduce some performance overhead. We do not suggest eliminating \texttt{PREFETCHW} since it is important for improving write performance.

Similar to prior cache attacks, our attacks also work by manipulating and monitoring cache states. Thus, defenses against prior cache attacks (e.g.,~\cite{bce,dawg,constant_time_guidance,timer_defense1,timer_defense2}) may also work on our attacks. For example, DAWG~\cite{dawg} allows replicated cache copies of the data shared across security domains: each domain gets their own copy of this data in cache. Thus, the attacker cannot monitor the coherence state changes from a victim who is in another domain, which would stop our attacks.

\section{Conclusion}
In this paper, we proposed a new cache eviction method as well as two new two cross-core cache side channel attacks that work with both inclusive and non-inclusive LLCs. One of the prefetch instructions on x86 processors, \texttt{PREFETCHW}, prepares the data for future writes by modifying the coherence state of the data. In this work, we made two important microarchitectural observations on \texttt{PREFETCHW}. First, it works on data with read-only permission. Second, its execution time is related to the coherence state of the target data. 
Given these observations, the coherence state modifications by \texttt{PREFETCHW} enable significant security vulnerabilities. Using \texttt{PREFETCHW}, we first built two covert channel attacks which have very high capacities. We also demonstrated that these high-capacity covert channels enable more powerful transient execution attacks. We then slightly modified the covert channel attacks to build two side channel attacks and showed that these attacks leaked information from real-world applications.

\section{Acknowledgement}
We thank the anonymous IEEE S\&P 2022 reviewers for their insightful feedback. We would also like to thank Daniel Weber for the help on the Meltdown implementation, and Daniel Gruss for his comments on the preprint. This work is supported in part by US National Science
Foundation \#1422331, \#1535755, \#1617071, \#1718080, \#1725657, \#1910413,
and \#2011146.


\bibliographystyle{ieeetr}
\bibliography{main}

\newpage
\section*{Appendix}

\setcounter{figure}{0}
\setcounter{table}{0}
\setcounter{lstlisting}{0}
\setcounter{footnote}{0}

\subsection{The Leakage Rate of Spectre v1}
\label{leakge_rate_spectre}

\begin{table}[!h]
\begin{center} \caption{The leakage rate of Spectre v1 when using Prefetch+Reload and Prefetch+Prefetch, respectively, normalized to Flush+Reload.}  
\setlength\arrayrulewidth{1.0pt}
\renewcommand{\arraystretch}{1.4}
    \setlength{\tabcolsep}{1.0mm}{ \begin{tabular}{lllll} 
    \hline
     & \multicolumn{2}{l}{Desktop processors} & \multicolumn{2}{l}{Server processors} \\
     \cline{2-5}
     Model & \vtop{\hbox{\strut Core}\hbox{\strut i7-6700}\hbox{\strut (3.4 GHz)}}& \vtop{\hbox{\strut Core}\hbox{\strut i7-7700K}\hbox{\strut (4.2 GHz)}}& \vtop{\hbox{\strut Xeon Platinum}\hbox{\strut 8124}\hbox{\strut (3.0 GHz)}} & \vtop{\hbox{\strut Xeon Platinum}\hbox{\strut 8151}\hbox{\strut (3.4 GHz)}} \\ \hline
     
     Prefetch+Reload & 1.61 & 2.40  & 1.57 & 1.64\\ 
     Prefetch+Prefetch & 3.02 & 3.94 & 2.01 & 2.08 \\ 
     \hline
\end{tabular}} \end{center}  
\end{table}

\subsection{The Meltdown Gadget}
\label{meltdown_gadget}

\begin{figure}[h]
\begin{minipage}{0.99\columnwidth}

\begin{lstlisting}[style=bb]
#define encode(x, b) ((((x) >> (b * 8)) & 0xff))
#define SPACING 4096
char mem[8][SPACING * 256];

uint64_t secret = *(uint64_t*)secret_addr;
memaccess(mem[0] + encode(secret, 0) * SPACING);
memaccess(mem[1] + encode(secret, 1) * SPACING);
memaccess(mem[2] + encode(secret, 2) * SPACING);
memaccess(mem[3] + encode(secret, 3) * SPACING);
memaccess(mem[4] + encode(secret, 4) * SPACING);
memaccess(mem[5] + encode(secret, 5) * SPACING);
memaccess(mem[6] + encode(secret, 6) * SPACING);
memaccess(mem[7] + encode(secret, 7) * SPACING);


\end{lstlisting}
\captionof{lstlisting}{The example Meltdown gadget where an access to the 64-secret is followed by eight secret encoding operations.}
\end{minipage}
\end{figure}

\end{document}